\documentclass{aa}  

\usepackage{graphicx}
\usepackage{txfonts}
\usepackage{url}
\usepackage{bm}

\begin{document}

   \title{Running with the bulls}
   \subtitle{The frequency of star-disc encounters in the Taurus star forming region}
   \authorrunning{Winter et al.}
   \author{Andrew J. Winter,
          \inst{1}\fnmsep\thanks{email: \email{andrew.winter@oca.eu}}
          Myriam Benisty,\inst{1, 2}
          Linling Shuai,\inst{3, 4}
          Gaspard D\^{u}chene,\inst{2, 5}
          Nicol\'{a}s Cuello,\inst{2}
          Rossella Anania,\inst{6}
          Corentin Cadiou,\inst{7}
          Isabelle Joncour\inst{2}
          }

   \institute{Universit{\'e} C{\^o}te d'Azur, Observatoire de la C{\^o}te d'Azur, CNRS, Laboratoire Lagrange, 06300 Nice, France
         \and
         Univ. Grenoble Alpes, CNRS, IPAG, 38000 Grenoble, France
         \and
          Astronomy Department, University of Michigan, Ann Arbor, MI 48109, USA
         \and
          Department of Astronomy, Xiamen University, 1 Zengcuoan West Road, Xiamen, Fujian 361005, China
         \and
         Astronomy Department, University of California Berkeley, Berkeley CA 94720-3411, USA  
         \and
        Dipartimento di Fisica ``Aldo Pontremoli'', Universita degli Studi di Milano, via Celoria 16, Milano, 20133, Italy
        \and
        Lund Observatory, Division of Astrophysics, Department of Physics, Lund University, Box 43, SE-221 00 Lund, Sweden
        }

   \date{Received September 15, 1996; accepted March 16, 1997}

% \abstract{}{}{}{}{} 
% 5 {} token are mandatory
 
  \abstract
  % context heading (optional)
  % {} leave it empty if necessary  
  {Stars and planets form in regions of enhanced stellar density, subjecting protoplanetary discs to gravitational perturbations from neighbouring stars. Observations in the Taurus star-forming have uncovered evidence of at least three recent, star-disc encounters that have truncated discs (HV/DO Tau, RW Aurigae, UX Tau), raising questions about the frequency of such events.}
   {We aim to assess the probability of observing truncating star-disc encounters in Taurus. }
   {We generate a physically motivated dynamical model including binaries and spatial-kinematic substructure to follow the historical dynamical evolution and stellar encounters in the Taurus star forming region. We track the star-disc encounters and outer disc radius evolution over the lifetime of Taurus.}
   {A quarter of discs are truncated below 30~au by dynamical encounters, but this truncation mostly occurs in binaries over the course of a few orbital periods, on a time-scale $\lesssim 0.1$~Myr. Nonetheless, some truncating encounters still occur up to the present age of Taurus. Strongly truncating encounters (ejecting $\gtrsim 10$~percent of the disc mass) occur at a rate $\sim 10$~Myr$^{-1}$, sufficient to explain the encounter between HV and DO Tau $\sim 0.1$~Myr ago. If encounters that eject only $\sim 1$~percent of the disc mass are responsible for RW Aurigae and UX Tau, then they are also expected with encounter rate $\Gamma_\mathrm{enc} \sim 100{-}200$~Myr$^{-1}$. However, the observed sample of recent encounters is probably incomplete, since these examples occurred in systems that are not consistent with random drawing from the mass function. One more observed example would statistically imply additional physics, such as replenishment of the outer disc material.}
   {The marginal consistency of the frequency of observed recent star-disc encounters with theoretical expectations underlines the value of future large surveys searching for external structures associated with recent encounters. The outcome of such a survey offers a highly constraining, novel probe of protoplanetary disc physics. }


    \maketitle
%
%________________________________________________________________

\section{Introduction}

Planet formation proceeds in a `protoplanetary disc' of dust and gas over a time-scale of $\sim 3$~Myr \citep[e.g.][]{Haisch01}. 
During this time, nascent planetary systems typically inhabit star forming regions with a local stellar density that far exceeds the average in the galactic neighbourhood \citep[e.g.][]{Lada03}. 
Neighbouring stars in these regions can feedback on the planet formation process in a variety of ways. 
These may include: external irradiation driving thermal winds \citep[][and references therein]{WinterHaworth22}, chemical enrichment \citep{Bastian13, Lichtenberg16, Parker23_enrichment}, late-stage gas in-fall \citep{Dullemond19, Kuffmeier20, Kuffmeier21, Kuffmeier23} and star-disc encounters \citep[][]{Cuello19, Cuello20, Cuello23}.
While all of these processes are of great interest for understanding the diversity of the observed exoplanet population, in this work we focus on the latter phenomenon. 
`Star-disc encounters' refer to gravitational perturbations experienced by a protoplanetary disc during the close passage of a neighbouring star.

Various phenomena such as stellar accretion outbursts \citep{Pfalzner08, Forgan10, Vorobyov21, Dong22} and spiral arms in protoplanetary discs \citep[e.g.][]{deRosa19} and free-floating planets \citep[e.g.][]{Vorobyov17} -- possibly including enigmatic binary planet-mass objects in the Orion Nebula cluster \citep{Pearson23, Wang23, PortegiesZwart23} -- may all be feasibly attributed to stellar encounters. There remain alternative explanations; for example, spiral arms may be produced by gravitational instability \citep[e.g.][]{Douglas13, Meru17, Baehr21} or companion stars/brown dwarfs/planets \citep[e.g.][]{Dong15, Ren20}. A search for flyby candidates that may have generated spiral arms by \citet{Shuai22} revealed no evidence that nearby stars recently had a sufficiently close approach. However, such an effort is challenging due to proper motion uncertainties and incompleteness of reliable measurements; particularly if a perturber is very low mass or is itself a binary. To assess the role of star-disc encounters for protoplanetary disc evolution, we must estimate the rate of encounters in dynamically evolving star forming regions.  

The rate of stellar encounters, and their role for planet formation, is dependent on the local stellar density of the star forming region.
The stellar mass density $ \rho_* $ of star forming regions can vary dramatically in the range $1\, M_\odot~$pc$^{-3} \lesssim \rho_*\lesssim 10^6 \, M_\odot~$pc$^{-3}$. 
The latter limit represents the most extreme stellar densities in the galaxy, such as in the cores of globular clusters that remain bound for a Hubble time  \citep[e.g.][]{Krumholz19}. 
Intermediate densities $\rho_*\sim  10^3 M_\odot\,\mathrm{pc}^{-3}$ are typical in the cores of open clusters that may survive against galactic tides over $100$~Myr or Gyr time-scales. 
However, possibly the most common environments in which stars and planets form are the low density regions that produce loose `associations', which are globally unbound against galactic tides. 
The average density in such regions has been assumed to be too low to induce frequent tidal encounters throughout the disc lifetime \citep[e.g.][]{Winter18b}.

Taurus is an example of a (globally) low density region, which does however host convincing evidence of at least three recent close star-disc encounters. 
These cases are RW Aurigae \citep{Cabrit06, Dai15, Rodriguez18}, HV and DO Tau \citep{Howard13, Winter18c} and UX Tau \citep{Zapata20, Menard20}.
Each of these systems exhibits significant external structure which appears to be explained by recent flybys, sufficiently close such that they were capable of unbinding disc material. 
In an unstructured star forming region at the average density of Taurus, the probability of a one-off random encounter over the disc lifetime is vanishingly small.
This puzzle may appear to be partially resolved for some cases if the system is bound, such that close encounters occur periodically over the binary orbit.
However, in isolation this is not a sufficient explanation. While repeated encounters in a multiple system can drive spiral arms in the disc \citep[e.g.][]{Alaguero24}, the disc should also be rapidly truncated such that on subsequent close approaches the gravitational pertubation no longer produces large extended tidal tails \citep[e.g.][]{Menard20}. 
It is therefore necessary to perturb star-disc systems in Taurus stochastically, and do so over its $\sim 1{-}3$~Myr lifetime. In this manuscript, we aim to answer the question: \textit{do we expect sufficient star-disc encounters in Taurus to explain the frequency of observed recent encounters?}

The role of stellar encounters has been the focus of numerous studies focusing both on the evolution of protoplanetary discs \citep[e.g.][]{Clarke93, Ostriker94, Pfalzner05, Winter18} and mature planetary systems \citep[e.g.][]{Spurzem09, Shara16, Winter22, Li24}. Studies using N-body simulations often perform parameter studies or implement scaling relations to model star forming regions. For example, one approach is to adopt the putative mass-radius (e.g. total stellar mass $M_\mathrm{c}$, half-mass radius $R_\mathrm{c}$) relationship for star forming regions \citep[e.g. $R_\mathrm{c} \propto M_\mathrm{c}^{1/2}$ --][]{Adams06}. However, the nature of this relationship varies depending on the sample/definition of mass and radius \citep{Pfalzner21}, and exhibits significant scatter. 
Mature/massive star clusters also do not appear to follow this relationship \citep{Krumholz19}, and particularly for low mass/density regions it is challenging to unambiguously define an individual star forming region.
In addition, even within a well-defined `individual' star forming region, internal structure has been shown to have a strong influence on the role of encounters \citep[e.g.][]{Cra13, Parker23}. 
Assessing encounter rates therefore requires quantitatively matching present day position-velocity structure. Yet it is not clear how well the widely adopted method for generating fractal initial conditions, sampling hierarchical boxes \citep{Cra13}, reproduces the observed spatial and kinematic structure in star forming regions.

 Structure in giant molecular clouds is set by the turbulent fragmentation, from which the power-spectrum and Mach number determine the mass density distribution \citep[e.g.][]{VazquezSemadeni94, Padoan97}.
An ideal approach to studying the role of dynamical encounters in young star forming regions is therefore to model the star formation process directly through hydrodynamic simulations. 
Such simulations have shown that encounters are common during early disc evolution, possibly determining the initial distribution of protoplanetary disc radii \citep{Bate18}. 
However, these experiments are computationally expensive, following star formation only for $\sim 0.1$~Myr time-scales.
This makes parameter studies or tailored modelling of individual star forming regions in this way impracticable.

Here we present a complementary approach to the above works, targeted at generating N-body initial conditions tailored to match young star forming regions. We achieve this by simulating a Taurus-like star forming region, with physically and empirically motivated initial conditions, including binaries. We draw stellar positions and velocities from an empirically constrained power spectrum, reflecting how turbulent energy is distributed across different scales. Our main aim is to develop a dynamical model to track the history of stellar encounters in the Taurus star forming regions. In doing so, we will assess whether the rate of disc-truncating star-disc encounters in Taurus is sufficient to produce the three known examples: HV/DO Tau, RW Aurigae and UX Tau. This goal requires ensuring that our dynamical model closely reproduces the present day spatial and kinematic structure in Taurus. We therefore review these structural properties in Section~\ref{sec:Taurus_review}, which we then adopt for benchmarking our dynamical model. We discuss our approach for initialising initial conditions and dynamically evolving the model in Section~\ref{sec:method}, in which we also draw comparisons with the structure metrics introduced in Section~\ref{sec:Taurus_review}. We discuss the rate of truncating encounters over time in our simulation in Section~\ref{sec:results}, quantifying the degree to which the observed examples of recent star-disc encounters are statistically expected. We summarise our conclusions in Section~\ref{sec:conclusions}.

\section{Properties of the Taurus star forming region }

\label{sec:Taurus_review}

\subsection{Aim}

An empirically motivated dynamical model for Taurus requires accurately quantifying kinematic substructure. If we had arbitrarily accurate and complete data for the 3D positions and velocities of all the stars in Taurus, then it would be trivial to use these data to generate N-body initial conditions to compute the future evolution of Taurus. However, as discussed in Section~\ref{sec:data}, while we have a fairly complete census of stars in terms of their projected spatial distribution, kinematic data is far more limited. Given the effects of multiplicity and extinction, parallax measurements also typically have associated uncertainties that make them impractical for use in setting initial conditions. In addition, we are interested in quantifying the frequency of encounters in the past, while we do not have direct measurements of the early spatial-kinematic stellar configuration in Taurus. For these reasons, we require metrics that characterise kinematic substructure. These metrics both guide our choice of initial conditions and offer a benchmark comparison for our models.

In the following, we first review the data for the stellar population in Taurus (Section~\ref{sec:data}). We then discuss how we quantify spatial structure in Section~\ref{sec:spatial_struct} and kinematic structure in Section~\ref{sec:vstruct_taurus}.

\subsection{Data}
\label{sec:data}
We use the census of Taurus members by \citet{Luhman23}, with astrometric data from \textit{Gaia} DR3 \citep{Gaia16, Gaia23, Babuxiaux23}. This catalogue contains 532 members, with a high degree of completeness for spectral types earlier than M6–M7. {When we consider proper motion differences between neighbours (Section~\ref{sec:vstruct_taurus}),} we restrict the sample to the 271 that have a reliable astrometric solution by the canonical criteria that the Renormalised Unit Weight Error (RUWE\footnote{\url{https://www.cosmos.esa.int/web/gaia/public-dpac-documents}}) is smaller than $1.4$ in \textit{Gaia} DR3. 

We cannot use the observed 3D positions directly to generate initial conditions for several reasons. Most obviously, the sample for which we have parallax is incomplete (417/532), and the typical uncertainties on the parallax correspond to a few~pc. This means that small scale structure is not resolvable. We therefore generate our initial conditions parametrically, closely comparing with the observed structure in Taurus. We consider only the projected separations between stars when inferring the spatial and velocity structure of the region.

\subsection{Spatial substructure}
\label{sec:spatial_struct}

We aim to understand the role of close neighbours on disc evolution. A sensible metric to quantify structure is therefore the normalised pair separation function, which can be defined in two dimensions $\hat{\Sigma}_\mathrm{pairs}(\Delta R)$. This is the averaged surface density of neighbours for any given star, normalised to unity when integrated over $2\pi \Delta R \, \mathrm{d}\Delta R$. This surface density evolves over time, and we will therefore use this metric to ensure that we capture the time at which the dynamical state in our simulation is similar to that in Taurus.

Two related but complementary metrics are the one- and two-point correlation functions, $\Psi(\Delta R) $ and $\xi(\Delta R)$ respectively. These metrics have been applied by \citet{Joncour17} and \citet{Joncour18} to study the structure in Taurus. They are broadly defined as the excess of pairs with a given separation compared to a random, uniformly distributed population of stars in the same area. The one-point correlation $\Psi$ is an excess of \textit{nearest} pairs, while the two-point correlation represents the excess of \textit{all} pairs. Of particular relevance for this work, \citet{Joncour17} applied the one-point correlation function to demonstrate that approximately $40$~percent of stars in Taurus are in ultra-wide binaries, with separations in the range $1.6\times 10^3 - 5\times 10^4$~au. We will use this inference to motivate our initial conditions, and the one- and two-point correlation functions to validate our model \textit{a posteriori}. In particular, from the one-point correlation function \citet{Joncour17} find a region of `inhibition' (a smaller number of pairs than expected from random sampling) between $\sim 0.1^\circ$ and $\sim 0.5^\circ$, while up to $\sim 0.2^\circ$, $\Psi \propto \Delta R^{-1.5}$. For the two-point correlation function, $\xi>1$ out to $\Delta R \approx 2^\circ$.

\subsection{Velocity substructure}
\label{sec:vstruct_taurus}

Stellar velocities are inherited from the velocities of the material from which they form, and are thus dependent on the kinematic structure of the parental molecular cloud. This velocity structure in a turbulent medium is driven by interacting waves that generate an energy cascade that is described by an energy spectrum $E(k)$. Across a wide range of length scales $\lambda$, it can be approximated by a power-law \citep[e.g.][]{Elmegreen04}:
\begin{equation}
\label{eq:energy_spectrum}
    E(k) = P(k)  k^2 \sim \frac{\sigma_{v,k}^2}{k} \propto k^{-\beta} \propto \lambda^\beta,
\end{equation}where $P(k)$ is the (three dimensional) power-spectrum, $k$ is the wavenumber, $\sigma_{v,k}$ is the characteristic velocity for $k$. 

\begin{figure}
    \centering
    \includegraphics[width=\columnwidth]{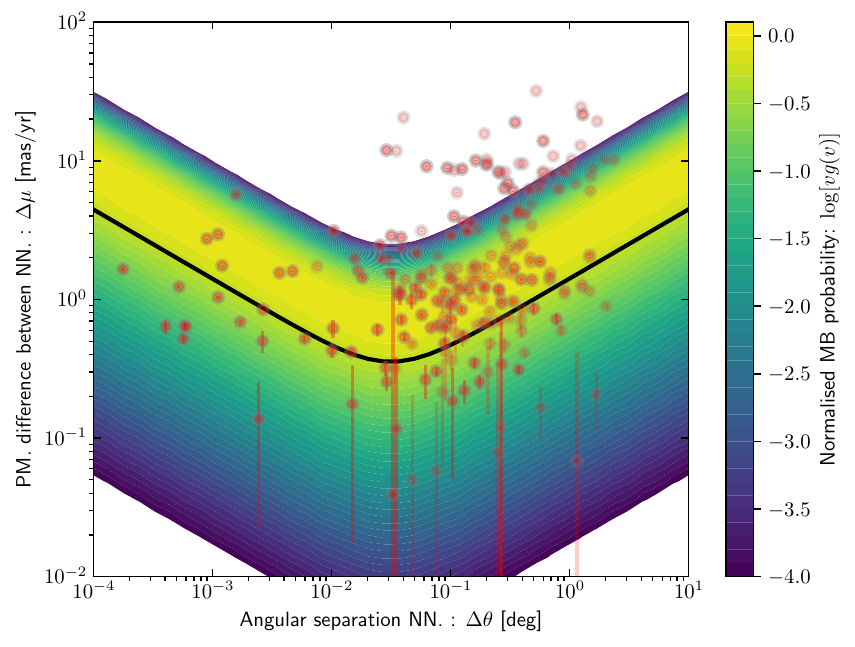}
    \caption{Proper motion difference as a function of angular separation on the plane of the sky for nearest neighbours in the \citet{Luhman23} sample for Taurus. Red data points are the observed sample, with uncertainties propagated from \textit{Gaia} uncertainties. The black line shows the the velocity dispersion as a function of separation we adopt for our model, with a Keplerian component that dominates at small separations. The transition between the two power-laws is the break between the binary and individual systems ($\sim 5\times 10^4$~au, comparable to the galactic tidal radius for typical stars). The colour bar shows the corresponding Maxwell-Boltzmann distribution, normalised for each separation. See text for details.  }
    \label{fig:dvNN_Taurus}
\end{figure}

A range of $\beta$ values in the interstellar medium (ISM) have been inferred observationally. These range from the original estimate for the size-linewidth relation for molecular clouds yielding $\beta \approx 1.76$ \citep{Larson81} to larger $\beta \approx 2.3$ \citep[][for example]{Munch58, Heyer04}. For weakly compressible turbulence, the energy spectrum for density and velocity follow the same power-law. \citet{Qian18} find $\beta\approx 2$ on small scales $\lesssim 2$~pc (suggestive of compressible turbulence) and $\beta \approx 5/3$ in the range $\sim 5{-}10$~pc \citep[suggestive of incompressible turbulence, see also][]{Brunt10}. \citet{Ha22} estimated $\beta \approx 1.7$ from the stellar population and $\beta \approx 1.8$ from H$\alpha$, both on $\sim 10$~pc scales. 

For our purposes we adopt $\beta=2$, appropriate for supersonic, rapidly cooling turbulence \citep{Burgers39}. The velocity dispersion relation we adopt is then:
\begin{equation}
\label{eq:sigv_fitfunc}
    \sigma_v(\Delta r) = \sigma_{v,0} \left(\frac{\Delta r}{\Delta r_0}\right)^{0.5},
\end{equation}where we choose normalisation constants $\sigma_{v,0} = 0.6$~km~s$^{-1}$ for $\Delta r_0 = 1 $~pc. This yields a velocity dispersion similar to that empirically inferred for the youngest stars on galactic scales in the solar neighbourhood \citep[e.g.][]{Holmberg09}, while more importantly matching the dispersion in Taurus down to binary length scales. In Figure~\ref{fig:dvNN_Taurus}, we show the distribution of proper motion differences for nearest neighbours as a function of their separation in Taurus. We also show the Maxwell-Boltzmann distribution:
\begin{equation}
g_{N_\mathrm{D}}(\Delta v | \sigma_v) =  \frac{2 \Delta v^{N_\mathrm{D}-1}}{\Gamma(N_\mathrm{D}/2)} \left(\frac{1}{4\sigma_{v} ^2}\right)^{{N_\mathrm{D}}/2} \exp\left( \frac{-\Delta v^2}{4 \sigma_v^2}\right)
\end{equation}in two dimensions ($N_\mathrm{D} = 2$) for equation~\ref{eq:sigv_fitfunc} with a Keplerian component appropriate for a star of mass $m_* = 0.5 \, M_\odot$ added in quadrature. This shows that the majority of neighbours follow the expected separation-velocity relation. Neighbours with a much larger relative velocity may have larger physical than projected separations. We will apply the size-velocity relation given by equation~\ref{eq:sigv_fitfunc} to generate stellar velocities, and compare the appropriate Maxwell-Boltzmann distribution to the nearest neighbour relative velocities we generate in our model, both initially and after dynamical evolution.

\section{Numerical method}
\label{sec:method}
\subsection{Overview}

Our approach for simulating a Taurus-like star forming region is to use a physically and empirically motivated set of initial conditions, including binary systems. We then benchmark our model against the spatial-kinematic properties of the present day Taurus to ensure that the simulation reflects the observed dynamical state. With this model, we are able to extract encounters which we can convert into the protoplanetary disc radii evolution using analytic formulae fit to numerical experiments for truncating encounters. 

In the remainder of this section, we detail this process. First, we discuss our method for generating physically motivated initial conditions, starting with a gas density distribution in Section~\ref{sec:gas_density}, which we convert to a stellar density distribution in Section~\ref{sec:single_stars}. We then discuss implementing empirically motivated binaries in Section~\ref{sec:binaries} and the appropriate velocity substructure in Section~\ref{sec:velocities}. We validate our dynamical model with respect to the observed dynamical state in Taurus in Section~\ref{sec:validate}. Finally, we discuss our approach for extracting close stellar encounters from the simulation in Section~\ref{sec:encounters} and the resultant disc evolution in Section~\ref{sec:rtrunc}.

\subsection{Nbody code}
\label{sec:nbodycode}
Throughout this work, we use \textsc{NBODY6++} \citep{Aarseth03,Spurzem99, Wang15} to integrate the stars and binary systems under gravity for $3$~Myr. Given our interest in the short term evolution of a low mass star forming region, we do not include stellar evolution. Nor do we include tidal binary circularisation, or an external potential. To capture the role of several neighbours, we insist on a short time–step factor for the irregular force polynomial $5\times 10^{-3}$. The parameter adjustment time in N-body units is $10^{-4}$ in code time-units that are $3.22$~Myr, {at which the regularisation parameters are updated}. The parameter and initial condition file for our fiducial model are available online.

\subsection{Lognormal gas density field}
\label{sec:gas_density}
We aim to initially generate a physically motivated gas density distribution, from which to draw our stellar population. Numerical experiments have shown that the mass density $\rho$ of isothermal turbulent flows is well approximated by a lognormal distribution \citep[e.g.][]{VazquezSemadeni94, Nordlund99, Ostriker99}. In order to generate a lognormal gas distribution, we must first generate a Gaussian density field over a 3D grid. Generating a density field that we can use to produce initial conditions for the N-body simulation is not trivial, since a Gaussian random field may exhibit peaks close to the edges of the grid. These peaks would be the site of high stellar density \citep[or `NESTS' --][]{Joncour18}, and may be partially excluded by the grid. Indeed, the centre of the grid can represent a `void' (underdensity) in the Gaussian field, which would undermine our goal to explore the interactions within and between stars in overdensities. Since we also need a grid that covers a large dynamical range (from the size scale of Taurus, down to the wide-binary scale), it is not practicable to simply choose a subset of the complete Gaussian random field with a very high resolution. It is therefore useful to apply an approach that ensures that we can centre the grid on an overdensity. 

While we could attempt to do this by multiplying the density field by a centrally concentrated profile (such as a 3D Gaussian or Plummer profile), this would fundamentally change the nature of the density field and corresponding power spectrum. To avoid this we adopt a two-stage process, inspired by cosmological zoom-in simulations and applying the publicly available \textsc{genetIC}\footnote{\url{https://github.com/pynbody/genetIC}} code \citep{Stopyra20,Stopyra21}. This code generates a zoom-in or splice \citep{Cadiou21} self-consistently within a surrounding Gaussian random field by Fourier-space filtering. The net effect is that we are able to generate an initial, coarse Gaussian field, and then zoom in on an overdensity that represents the centre of our model for Taurus.

\begin{figure}
    \centering
    \includegraphics[width=\columnwidth]{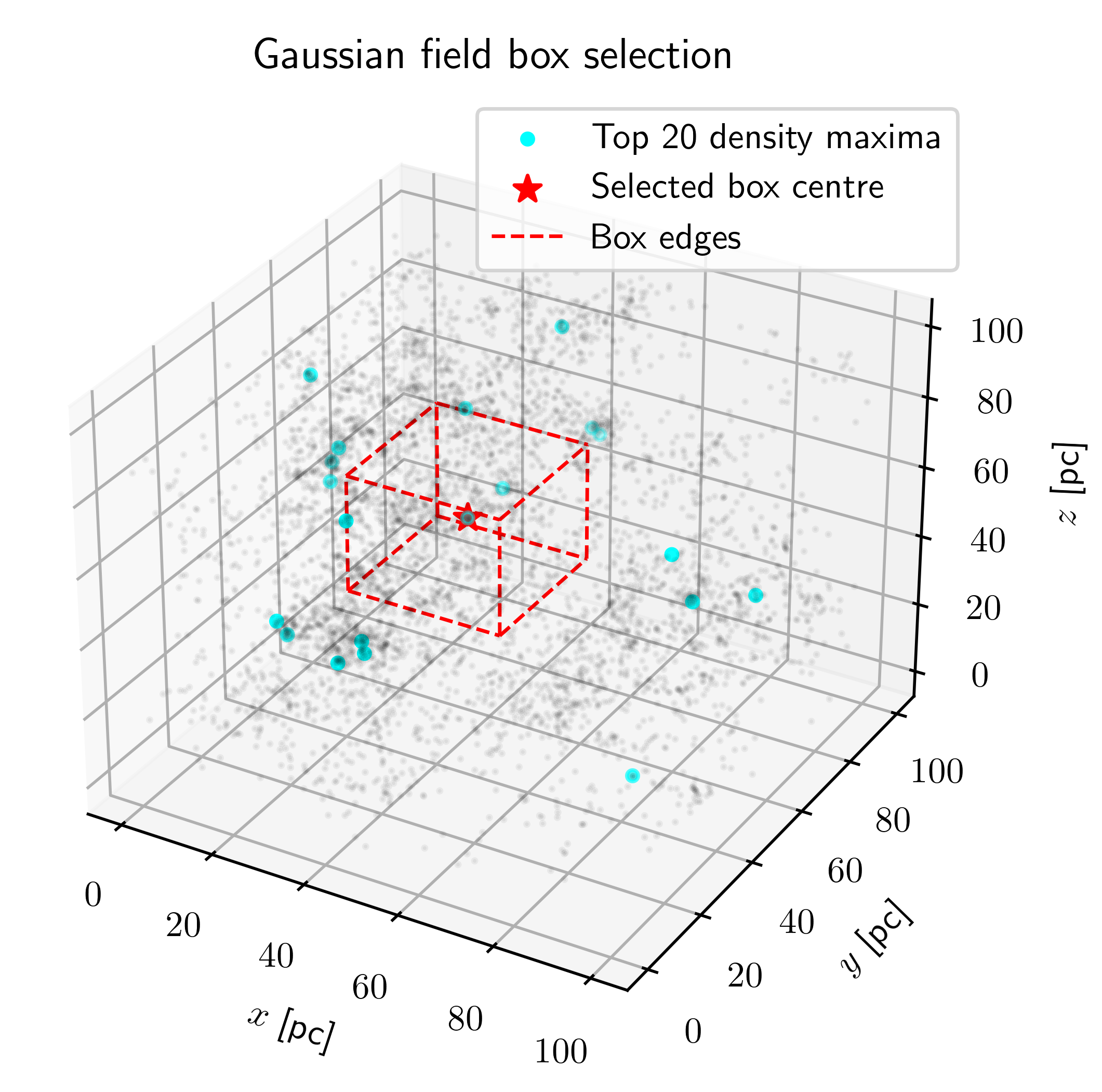}
    \caption{Illustration of how we select a box from which to generate a lognormal density field from a larger scale, low resolution lognormal density field. We illustrate the underlying density grid by drawing faint black points with density proportional to the local density field. Cyan points indicate our selection of the top 20 density maxima from the density field. The red star indicates the position of the adopted box centre. The dashed red lines indicate the boundary of the new zoom-in box. }
    \label{fig:box_selection}
\end{figure}

Our approach for this is represented graphically in Figure~\ref{fig:box_selection}. We first generate a Gaussian random field over a $256^3$ grid with side length $100$~pc. We adopt a power spectrum index of $-3$. We do not quantitatively fit for this index, but select \textit{a posteriori}  between indices $-2$, $-2.5$, $-3$ and $-4$, finding that $-3$ reproduces a stellar density structure that matches the observed structure. Empirically, this index is comparable to that inferred in Taurus by \citet{Brunt10}. 

From the large scale density field, we then locate local maxima over $5$~pc length scales using the \texttt{ndimage.maximum\_filter} algorithm from \textsc{Scipy} \citep{Virtanen20scipy}. We choose the 20 greatest local maxima, and select the closest to the centre of the grid, ensuring the zoom-in grid will be within the range of the initial grid. We then use \textsc{genetIC} to generate a new Gaussian density field on a grid that is smaller by a factor three (box side length $33$~pc), and with $1024^3$ grid cells. This defines a field which has a high density centre, and a resolution of $0.032$~pc, or $\sim 7000$~au, down to length scales comparable to wide binary separations. Finally, we enforce a dispersion $\sigma_{\ln \rho} = \ln \left(1 + 0.25  \langle v^2 \rangle /c_\mathrm{s}^2\right) = 5$ on the logarithmic density field. This is approximately the expected dispersion given a sound speed $c_\mathrm{s}= 0.2$~km~s$^{-1}$ and the root mean square velocity $\langle v^2 \rangle^{1/2}=  \sqrt{3} \sigma_v \sim 5$~km~s$^{-1}$ for $\Delta r \sim 30$~pc, according to equation~\ref{eq:sigv_fitfunc}. 

\subsection{Stellar spatial distribution}
\label{sec:single_stars}
We now wish to draw a stellar spatial distribution from our underlying gas density profile. To do this, we must consider which regions of the cloud are able to rapidly collapse to form stars. We consider the time-scale for this collapse from rest:
\begin{equation}
    \tau_\mathrm{ff} = 1.52 \left( \frac{\rho}{100 \, M_\odot \,\mathrm{pc}^{-3}}\right)^{-1/2} \, \mathrm{Myr}.
\end{equation}To infer a free-fall time across our grid, we require an absolute density scale, for which we assume a total mass within the box of $10^4 \, M_\odot$, corresponding to the approximate total gas mass in the Taurus complex \citep{Goldsmith08}. We then impose a constraint that grid cells must have $\tau_\mathrm{ff} < 1$~Myr in order to host a star, comparable to the empirical age dispersion \cite{Luhman23}. We assign probabilities to these grid cells $p_* \propto \rho \tau_\mathrm{ff}^{-1} \propto \rho^{3/2}$ -- i.e. the probability of forming a star is proportional to the quantity of material multiplied by the rate at which that material is expected to collapse. For a selected grid cell, we then offset the location of the formed star such that it is placed randomly within the cell. We validate this choice of spatial distribution in our dynamical model \textit{a posteriori} in Section~\ref{sec:validate}. 

For the the total number of systems, we draw $N_\mathrm{sys}$ that yields a total number of stars $N_*$ that is broadly consistent with the findings of \citet{Luhman23}. This is not trivial, both because there is not a clear detection limit for that sample and because we include binaries stochastically from the initial population of potential primaries, as well as the ultrawide binary fraction inferred by \citet{Joncour17}. Further, it is not straightforward to interpret what fraction of binaries are resolved/detected by \citet{Luhman23}. {We perform a two stage process as follows in Section~\ref{sec:uwps} and~\ref{sec:binaries}.} 

\subsection{Ultrawide pairs}
\label{sec:uwps}
{First of all, we correct the total number of stars for a given fraction of ultrawide binaries $\mathcal{F}_{\mathrm{uwb}} =2/3$. We consider $400(1-\mathcal{F}_\mathrm{uwb}/2) = 267$ stars with masses $>0.08 \, M_\odot$ (half of these will end up in ultrawide binaries). We then add the fraction of brown dwarfs we expect below this limit from the initial mass function (IMF). }For this, we use the following IMF \citep[e.g.][]{Kro01}:
\begin{equation}
\label{eq:imf}
    \xi(m_*) \propto  \begin{cases}
        m_* ^ {-0.3} & \qquad 0.01 \, M_\odot \leq m_*<  0.08 \,M_\odot \\
        m_* ^ {-1.3} & \qquad 0.08 \, M_\odot \leq m_*<  1.3 \,M_\odot \\
        m_* ^ {-2.3}& \qquad 0.5 \, M_\odot \leq m_* < 1.3 \,M_\odot\\
        m_* ^ {-2.7} & \qquad  1.3 \, M_\odot\leq m_* \\
        0 &\qquad \mathrm{otherwise}
    \end{cases},
\end{equation}with normalisation constants such that $\xi$ is continuous and integrates to unity over all $m_*$. For each of these stars, we then draw an ultrawide companion with total probability $\mathcal{F}_{\mathrm{uwb}}$. This yields $660$ stars and sub-stellar objects in total, $38$~percent of which have masses $<0.08\, M_\odot$. We then draw the masses of all stars from the equation~\ref{eq:imf}. 

{The ultrawide pair fraction we adopt is somewhat higher than $\mathcal{F}_{\mathrm{uwb}} \approx 0.55$, inferred by \citet{Joncour17}; 186 of the 338 stars in their sample are in ultrawide pairs. However, some of these companions will separate during dynamic evolution. We also note that \citet{Joncour17} found evidence that members of ultrawide pairs are $\sim 15$~percent more likely to be themselves in a shorter period binary. However, given that this is a relatively minor correction to the binary fraction, and it is challenging to disentangle dynamical versus primordial origins for these statistics, we do not include this enhancement. While we do not forward model the initial fraction, we validate our choice by comparing the one-point correlation function in our model with that computed by \citet{Joncour17} in Section~\ref{sec:validate}.}

The separations for the ultrawide pairs is approximately log-uniform between $1.6\times 10^3$~au and $5\times 10^4$~au \citep[see Fig. 7 of][]{Joncour17}, and we therefore draw semi-major axes of these companions similarly. We assume a uniform eccentricity distribution up to $0.9$, and a random orientation (as described in Section~\ref{sec:binaries}).

\begin{figure}
    \centering
    \includegraphics[width=\columnwidth]{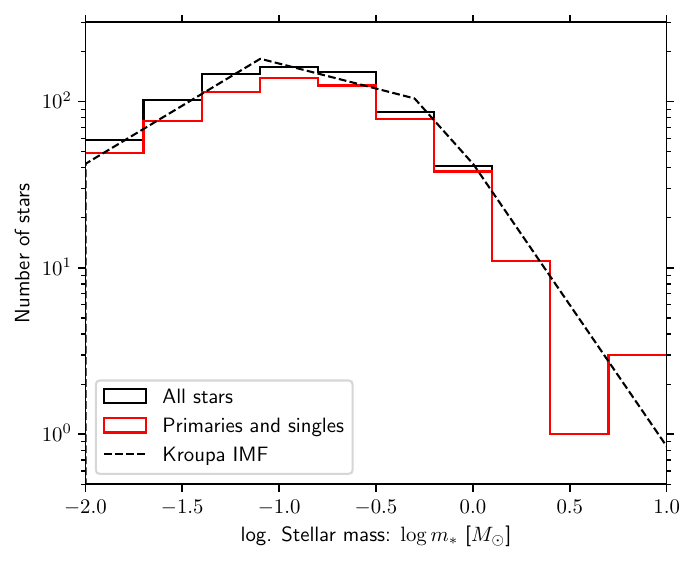}
    \caption{Histogram of stellar masses in our dynamical model for Taurus. The solid black shows all stars including binary companions, while the red line is just the primaries and single stars. The black dashed line corresponds to the number that would be expected in each bin from the \citet{Kro01} IMF (equation~\ref{eq:imf}). The bin sizes are similar to those adopted by \citet{Luhman04} in Figure 13, top panel.}
    \label{fig:model_mstar}
\end{figure}

\subsection{Binaries}
\label{sec:binaries}

We define binaries separately from `ultrawide pairs', described above and treated as single stars in terms of their mass-function.   We consider the entire population of stars (including each member of the ultrawide pairs) to be potential primary stars in a binary system, and add companions to a subset. 

The binary population we include is empirically motivated, based on the findings of \citet{Moe17}. In brief, we use the same functional form for the probability of each primary having a companion in each dex of orbital period space, modified in a number of ways. Firstly, we exclude binaries with period $P< 10^{5}$~days ($\lesssim 30$~au); such binaries are close enough to be largely uninfluenced by encounters, and in the context of protoplanetary discs may host circumbinary discs. At longer periods, we remove the exponential taper that \citet{Moe17} infers for $P > 10^{5.5}$~days, instead assuming a constant binary fraction per dex out to $P= 10^{7.7}$~days, approximately corresponding to the scale at which we transition to ultrawide binaries. The log-uniform distribution is consistent with the observed separations for the ultrawide pairs \citep[see Fig. 7 of][]{Joncour17}, and for wide binaries in the field \citep{Lepine07}. For all binaries, we draw eccentricities uniformly from $0{-}0.9$ \citep[e.g][and review by \citealt{Duchene13}]{Abt06}.

\begin{figure*}
    \centering
    \includegraphics[width=\textwidth]{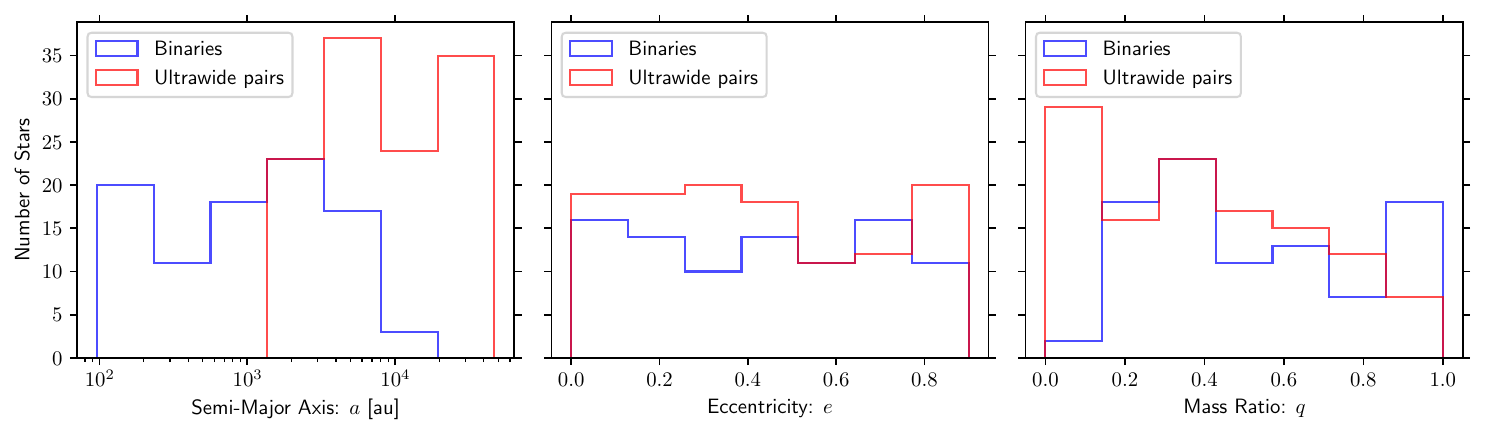}
    \caption{{Histogram showing the distribution of companions at a given semi-major axis (left) eccentricity (middle) and mass-ratio (right) for primary stars with masses $m_*>0.08\, M_\odot$. The red histogram is for those pairs we define as ultrawide pairs, while the blue shows the binaries.} }
    \label{fig:binstats}
\end{figure*}

Once we have randomly drawn the periods for the companions of our initial population, we draw the mass-ratio and eccentricity. For the ultrawide binaries, we draw masses from the IMF, as described in Section~\ref{sec:single_stars}. For the regular binaries, we draw mass-ratios $q$ in the range $0.1< q < 1$ from a probability density function $p(q)$. To construct $p$, we define two power-law regime with indices $\gamma_{\mathrm{small}}$ for $q<0.3$ and $\gamma_{\mathrm{large}}$ for $q\geq0.3$. Following \citet{Moe17}, we also include an excess probability of the star having a twin, where a twin is defined as having a mass ratio $1- \Delta_\mathrm{twin}< q \leq 1$. This is defined in practice by computing the quantity: 
\begin{equation}
    p_\mathrm{twin}' = \begin{cases} \frac{\mathcal{F}_\mathrm{twin} \int_{0.1} ^1 p' (\tilde{q}) \, \mathrm{d} \tilde{q}}{\Delta_\mathrm{twin} (1- \mathcal{F}_\mathrm{twin} ) } &\qquad 1-\Delta_\mathrm{twin} < q < 1 \\
    0 & \qquad \mathrm{otherwise}
    \end{cases},
\end{equation}where
\begin{equation}
    p' (q) = \begin{cases}
        q^{\gamma_{\mathrm{small}}} & \qquad 0.1\leq  q<0.3 \\
        \frac{0.3^{\gamma_{\mathrm{small}}}}{ 0.3^{\gamma_{\mathrm{large}}}} q^{\gamma_{\mathrm{large}}} & \qquad q\geq 0.3
    \end{cases}.
\end{equation}Then finally the probability density function for $q$ is:  
\begin{equation}
    p(q) = \frac{p' +p'_\mathrm{twin}}{\int_{0.1}^1 p'(\tilde{q}) +p'_\mathrm{twin}(\tilde{q}) \, \mathrm{d}\tilde{q} }.
\end{equation}Since we only consider $P > 10^{5}$~days and largely low mass stars, we have only two regimes for the power-law indices. For $P< 10^6$~days we have $\gamma_{\mathrm{small}} = 0.4$ and $\gamma_{\mathrm{large}} = -0.4$, and otherwise we have $\gamma_{\mathrm{small}} = 0.5$ and $\gamma_{\mathrm{large}} = -1.1$. We fix $\mathcal{F}_\mathrm{twin} = 0.1$, for $
\Delta_\mathrm{twin} = 0.05$. These values are consistent with observational constraints, although typical uncertainties are high for many of these values \citep{ElBadry19}. When we generate our initial population, we also exclude any companions that are generated that have masses below our lower IMF limit ($0.01 \, M_\odot$), although in practice this only influences brown dwarf primaries. The overall binary fraction is $\sim 30$~percent for binaries with orbital periods $>10^5$~days, which is broadly consistent with the field population~\citep[e.g.][]{Niu21}. {We show the statistics of our binary and ultrawide pair distribution for the stellar population in Figure~\ref{fig:binstats}.}

Finally, we generate the positions and velocities of the companion population by randomly sampling $\cos i$ (for inclination $i$), the argument of periapsis, longitude of ascending node and true anomaly uniformly over the appropriate ranges. We then compute the appropriate position and velocity vector of the companion with respect to the primary. We thus produce the initial binary population that we evolve dynamically during our simulation. 

\subsection{Mass function}

We have already described our stellar mass drawing procedure for single stars and binaries in Sections~\ref{sec:uwps} and~\ref{sec:binaries}. To validate our mass distribution including the binary population, we show the histogram of stellar masses in Figure~\ref{fig:model_mstar}. The bins used for this histogram are similar to those used by \citet[][top panel of their Figure~13]{Luhman04}. The number of approximately solar mass stars found by \citet{Luhman04} is $\sim 40$. This number is close to complete for Taurus, and is similar to our drawing from the IMF. More recently, \citet{Esplin19} found a peak around $m_*\sim 0.15$ and $\sim 25$ stars with $m_*\gtrsim 1 \, M_\odot$ across the whole of Taurus, again comparable to our IMF draw. The mass function for all the stars, including binary companions, is shown as the black histogram in Figure~\ref{fig:model_mstar}. Despite including a different mass function for companions, the mass function is not greatly altered from the \citet{Kro01} IMF we initially assume for the primary population. We therefore conclude that we have drawn a similar stellar population to that of Taurus. 

\subsection{Initial velocities}
\label{sec:velocities}

We must assign velocities to the primary stars we have generated, motivated by our discussion in Section~\ref{sec:vstruct_taurus}. We tackle this by generating velocities following a Gaussian process, such that velocities of stars that are close to each other are more highly correlated than those of stars at large separations. If $\Delta r < \Delta r_\mathrm{max} $, we can write the corresponding kernel or covariance function \citep[e.g. see equation~2.19 of][]{Rasmussen06}:
\begin{equation}
\label{eq:kfunc_lin}
    k(\bm{r},\bm{r'}) = k(\Delta r)  = \sigma_{v,\mathrm{max}}^2 - \sigma_v ^2 (\Delta r),
\end{equation}where $\sigma_{v,\mathrm{max}} = \sigma_{v}(\Delta r_\mathrm{max})$. By this definition, the covariance function is not well defined for large $\Delta r > \Delta r_\mathrm{max}$, where $k$ becomes negative. However, we  are free to choose an arbitrary large $\Delta r_\mathrm{max}$. In this case, we can also rewrite the covariance function: 
    
\begin{equation}
\label{eq:kfunc}
    k(\Delta r)  \approx \sigma_{v,\mathrm{max}}^2 \left( 1 - \mathcal{D}_r^2\right)
\end{equation}where we have defined:
\begin{equation}
    \mathcal{D}_r = \left[1- \exp\left(-\frac{\Delta r}{\Delta r_\mathrm{max}}\right) \right]^{0.5}.
\end{equation}

Equations~\ref{eq:kfunc_lin} and~\ref{eq:kfunc} are equivalent for large $\Delta r_\mathrm{max}/\Delta r\rightarrow \infty$, but at large separations equation~\ref{eq:kfunc} defines a maximal dispersion between stellar velocities. We choose equation~\ref{eq:kfunc} because it yields well defined covariance for any $\Delta r$. In practice, we anyway choose $\Delta r_\mathrm{max}=100$~pc so that our choice is not important.  In order to assign velocities given this kernel function, we define the covariance matrix $\bm{K} = [k_{ij}] = [k(\bm{r}_{i}, \bm{r}_{j})]$. We then perform a Cholesky decomposition $\bm{K} = \bm{L}\bm{L}^\mathrm{T}$, where \(\bm{L}\) is a lower triangular matrix. We define a vector $\bm{w}_{\alpha}$ which is a vector with a length corresponding to the number of stars for which we assign velocities. We draw each $w_{ j \alpha} \sim \mathcal{N}(0, 1)$ independently for each spatial dimension $\alpha=1, 2, 3$. We then define the velocity components for the stellar population $\bm{u}_{\alpha} = \bm{L}  \bm{w}_\alpha $.

\begin{figure}
    \centering
    \includegraphics[width=\columnwidth]{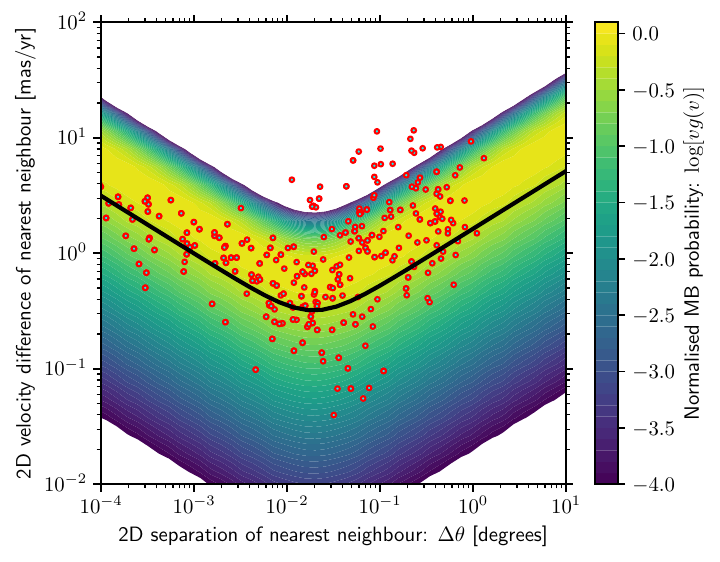}
    \caption{Data points show the proper motion difference versus angular separation (assuming a distance of $140$~pc) for nearest neighbours from the initial conditions in our model. The black line shows the one dimensional velocity dispersion as a function of separation and the colour bar shows the normalised Maxwell-Boltzmann distribution for each separation, as in Figure~\ref{fig:dvNN_Taurus}.}
    \label{fig:vdist_3D_IC}
\end{figure}

In Figure~\ref{fig:vdist_3D_IC} we show the velocity difference between different nearest neighbours in our model initial conditions, including binaries. It is clear from Figure~\ref{fig:vdist_3D_IC} that our synthetic stellar population (scatter points) have relative velocities that follow a similar size-velocity relation, $\sigma_v(\Delta r)$ as found in Section~\ref{sec:vstruct_taurus} -- cf. Figure~\ref{fig:dvNN_Taurus}. 

\subsection{Model validation}
\label{sec:validate}

\begin{figure}
    \centering
    \includegraphics[width=\columnwidth]{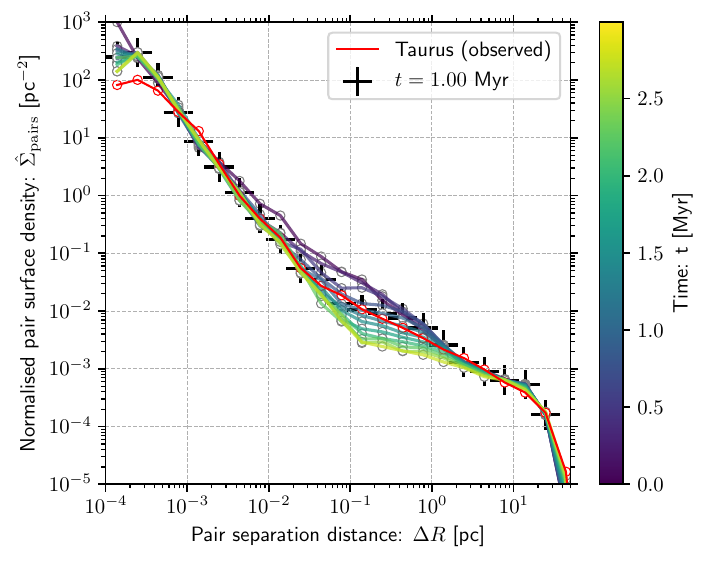}
    \caption{The normalised pair surface density $\hat{\Sigma}_\mathrm{pairs}$ as a function of projected separation $\Delta R$ in parsecs. The red line shows the observed surface density for Taurus from the \citet{Luhman23} census, which may be incomplete at the smallest separations. The coloured lines show the outcome from our model, at snapshot outputs indicated by the colour bar. {The $1$~Myr snapshot is also marked by black crosses, at which time the pair distribution function is in good agreement between observations and simulation.}}
    \label{fig:ic_pairs}
\end{figure}

We validate our model by analysing the structure metrics discussed in Section~\ref{sec:Taurus_review}. First, we consider the pairwise separation distribution, as shown in Figure~\ref{fig:ic_pairs}. The pair surface density profile $\hat{\Sigma}_\mathrm{pairs}$ shows an excess at very small projected separations ($\Delta R \lesssim 5 \times 10^{-4}$~pc), where the \citet{Luhman23} sample may be missing closer binaries. There is also a small excess around the binary transition at a few $10^{-2}$~pc. This excess is quickly lost as the system evolves, and the spatial structure is in good agreement with the observed population at $\sim 1$~Myr. We therefore adopt $1$~Myr as the `present day' in our simulation. 

\begin{figure}
    \centering
    \includegraphics[width=\columnwidth]{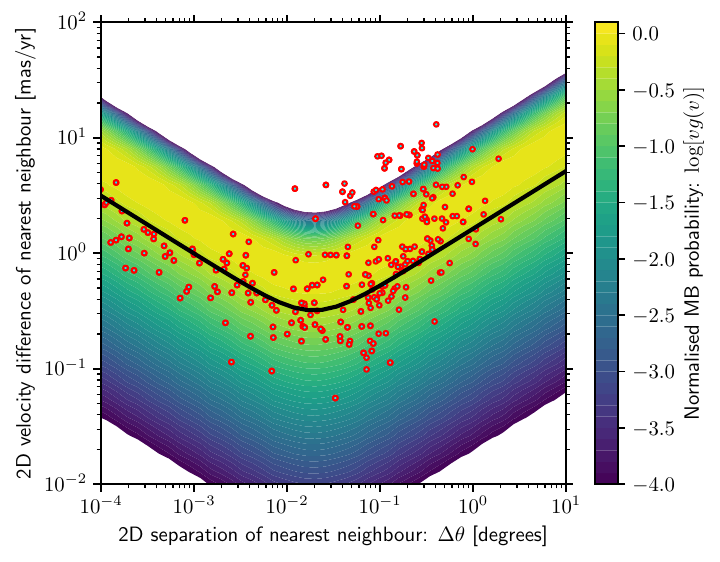}
    \caption{As in Figure~\ref{fig:vdist_3D_IC}, except for our simulation after $1$~Myr of evolution. }
    \label{fig:vdist_3D}
\end{figure}

At this time, the velocity structure (illustrated in Figure~\ref{fig:vdist_3D}) remains similar to the velocity structure we inferred in Taurus in Section~\ref{sec:vstruct_taurus}. When comparing Figure~\ref{fig:vdist_3D} to Figure~\ref{fig:dvNN_Taurus}, we note that there are some differences in how they are constructed. For example, close binaries are complete in our model, but not for those in Taurus. Indeed, in Taurus the sample is restricted to only stars with \textit{Gaia} proper motions, with all the biases that implies. Nonetheless, the correlation between velocity and separation remains broadly similar. The transition in nearest neighbour velocity difference from `binary' to `field' coincides with the change in the power-law surface density profile in Figure~\ref{fig:ic_pairs} for $\Delta r \sim 3\times  10^{-2}$~pc.

\begin{figure}
    \centering
    \includegraphics[width=\columnwidth]{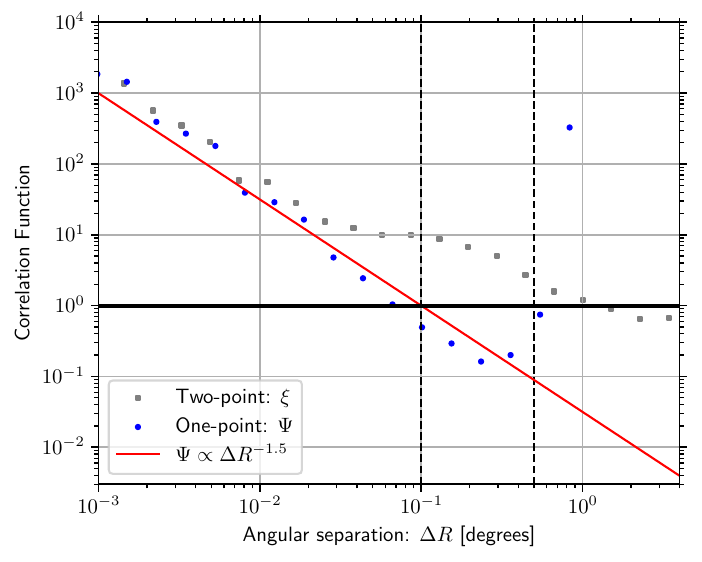}
    \caption{One-point ($\Psi$, blue points) and two-point ($\xi$, grey points) correlation functions computed from our model at $1$~Myr. The red line shows the $\Psi \propto \Delta R^{-1.5}$ relationship for $\Delta R < 0.2^\circ$ and the two dashed black lines enclose the region of inhibition (where $\Psi<1$), as inferred by \citet[][see their Figure 4]{Joncour17}. }
    \label{fig:corr_funcs}
\end{figure}

We also show the one- and two-point correlation functions in Figure~\ref{fig:corr_funcs}. These can be compared directly to Fig. 4 of \citet{Joncour17}. We find a similar functional form for both, again indicating that the structure in our model matches the physical structure of Taurus. We obtain a very similar region of `inhibition' in which $\Psi<1$ in the range of separations $0.1-0.5^\circ$. We find a similar power-law for $\Psi$ where $\Delta R < 0.2^\circ$. We also recover $\xi>1$ for $\Delta R \lesssim 1^\circ$, close to the result of \cite{Joncour17}. 

We conclude that our model, with initial conditions based on physical and empirical arguments, reproduces the observed dynamical state of Taurus at $1$~Myr. The age in our model of $1$~Myr is somewhat younger than the $1-3$~Myr typically estimated for Taurus \citep[e.g.][]{Luhman23}. If Taurus is typically older on average, this might suggest that the stellar distribution was initially more compact than the initial conditions in our simulation; although in this case we would also expect more rapid dynamical evolution. It is also possible that residual gas in the star forming region slows down the dispersal of structure. However, then it would be plausible that discs in the high density regions are replenished by accretion of this residual gas \citep[e.g.][]{Kuffmeier20}, blurring the lines of what `$t=0$' means for disc evolution. In truth, probably a mixture of these influences, as well as a finite period of star formation, somewhat influence the dynamical evolution of the region. Nonetheless, for the purpose of this work, we are satisfied that the current state of stars and discs in Taurus is well approximated by our simulation at $1$~Myr. In this context, we note that \textit{a posteriori} we find that the frequency of strong encounters does not change rapidly between $1-3$~Myr in our simulation (see Section~\ref{sec:encounter_freq}).  

\subsection{Tracking encounters}
\label{sec:encounters}
Due partly to our inclusion of binaries, interactions between stars are too complex to be easily identified with a simple criterion `on the fly' during the simulation. We have therefore chosen to analyse encounter properties by post-processing high frequency outputs. Specifically, we have 9299 snapshots from our simulation over $3$~Myr, corresponding to a time-step of $323$~years. To accurately extract the correct encounter properties, we then perform the following analysis for each star $i$:
\begin{enumerate}
    \item Identify the closest neighbour $j(t)$ for each time-step, spatially separated by vector $\Delta \bm{r}_{ij}$ and with velocity difference $\Delta \bm{v}_{ij}$.
    \item If the nearest neighbour $j$ to $i$ has a bound companion $k$ which is separated from $j$ by vector $\Delta \bm{r}_{jk}$ such that $|\Delta \bm{r}_{jk}|<0.1 |\Delta \bm{r}_{ij}|$, we consider $j$ and $k$ to be a single star with the combined mass and momentum of $j$ and $k$. In the following we will refer to star $j$ as the binary barycentre. 
    \item We search for a sign change in the vector $\Delta \bm{r}_{ij} \cdot \Delta \bm{v}_{ij}$ from negative (approaching) to positive (receding). We define snapshot $l$, at time $t_l$, for which $|\Delta \bm{r}_{ij}|$ is minimal between the two adjacent snapshots ($l$, $l+1$).
    \item We compute the analytic eccentricity $e$, closest approach distance $r_\mathrm{p}$, time of pericentre $t_\mathrm{p}$. If the predicted closest approach is not between $t_l$ and $t_{l+1}$, then this must be a non-hierarchical multiple interaction, occurring on a short time-scale ($<323$~years). In this case, we assume that the closest approach is at $t_l$ (with closest approach distance given by the separation at this time), although in practice this is rare.
\end{enumerate}

This procedure ensures that, despite the finite temporal resolution of our simulations, we are able to resolve {the majority of} encounters that are relevant to disc truncation. We limit the number of encounters to one per time-step, thus we do not count every closest approach for close binaries which have an orbital period $<323$~years. Although we could in principle include multiple encounters per time-step, these encounters anyway quickly truncate the disc on short time-scales. {While our method may also not be accurate particularly for chaotic multiple interactions in cases where multiple interactions occur on time-scales less than $323$~years, we show in Appendix~\ref{app:output_freq} that our results are not influenced if we increase the output frequency.}

\subsection{Disc truncation and initial radii}
\label{sec:rtrunc}

To compute the post-encounter disc radius, we use the analytic functions inferred by \citet{Winter18b}. These functions were established by fitting a scale-free, angle-averaged expression to numerical test particle simulations, depending on the ratio of the closest approach distance $r_\mathrm{p}$ to the outer disc radius $R_\mathrm{out}$, the eccentricity of the encounter $e$ and the mass-ratio of the perturber $q$. Since these fitting functions were inferred for unbound encounters, we will adopt $e=1$ for encounters with $e<1$. We expect this to be a reasonable approximation. For example, \citet{Manara19} fit an analytic functional form to the steady state truncation radius inferred from the numerical results of \citet{Artymowicz94}. For an equal mass system on a circular orbit, this estimate implies a truncation radius $\sim R_\mathrm{out} \approx 0.33 r_\mathrm{p}$, which is the same as the steady state truncation radius implied by our prescription \citep[see Figure 4 of][]{Winter18b}. In practice, the majority of encounters that strongly truncate the disc at late times also have $e\sim 1$. We also consider all discs which experience encounters with $r_\mathrm{p}< 10$~au to be `destroyed', making the assumption that the resultant compact disc may be rapidly accreted or photoevaporated \citep[e.g.][]{Clarke01}. 

Unless otherwise stated, we initialise all discs with
outer radii following \citep{Andrews20}:
\begin{equation}
\label{eq:mass_radius_relation}
    R_{\mathrm{out},0} = 250 \left( \frac{m_*}{1\, M_\odot} \right)^{0.9} \, \mathrm{au}.
\end{equation}While this is empirically motivated by observed outer radii as inferred from CO emission lines, this relationship has a large scatter (which may in part be driven by dynamical truncation). The measurements themselves also come with numerous caveats originating from systematic uncertainties in outer radius definition, optical depth and chemistry \citep[see e.g.][]{Miotello23}. However, with these caveats we will here assume that the relationship is exact. We also do not consider viscous expansion \citep[e.g.][]{Lyn74} or wind-driven contraction \citep[e.g.][]{Tabone22}. The discs therefore only evolve (shrink) as a result of star-disc encounters. 

\subsection{Caveats for the physical model}
\label{sec:caveats}
{There are two main caveats for the dynamical model we present here. The first is that, while we are interested in establishing the rate of strongly disc-truncating interactions at a given system age, this depends on there being a well-defined age of the system at large. More specifically, we implicitly assume that the spread in stellar ages is much smaller than the age of the system itself. This is almost certainly not the case, with the groups in Taurus having ages in the range $\sim 1{-}3$~Myr \citep{Luhman23}. However, our efforts are still valid in that a star-disc system will only undergo encounters with nearby neighbours, which we generally expect to have very similar ages. None of the existing examples of recent encounters have been suggested to be particularly young, but our conclusions can be reconsidered depending on the inferred ages of post-encounter systems. }

{The other major caveat is that we do not include the self-gravity of the ISM. Without including a fully hydrodynamical model, it is not possible to include this potential in a physical way. Our choice to ignore the gas potential is in a sense a similar assumption to the one discussed above, specifically that local star formation is completed instantly, and the gas is instantaneously dispersed (e.g. by stellar feedback). If instead a large quantity of residual gas remained within the individual star forming regions that comprise the Taurus complex, then groups of stars may remain bound for longer. This may in turn increase the local encounter rate at later times. It is not possible to capture this gas potential without either full hydrodynamics (making matching exactly the structural properties of Taurus impossible), or by developing a scheme for integrating a complex potential that allows several components that co-move with stellar groups (beyond the scope of this work). However, we do find agreement between the dynamical state (i.e. correlation functions) in our model and observations, which would suggest the role of ISM self-gravity on the population dynamics as a whole cannot be dramatic. While this nonetheless remains a shortcoming of this work, we discuss in Section~\ref{sec:add_physics} that if high density gas alters our results this would have other interesting physical implications.}

\section{Results and discussion}
\label{sec:results}

\subsection{Types of late-stage encounter}

\begin{figure*}
    \centering
    \includegraphics[width=\textwidth]{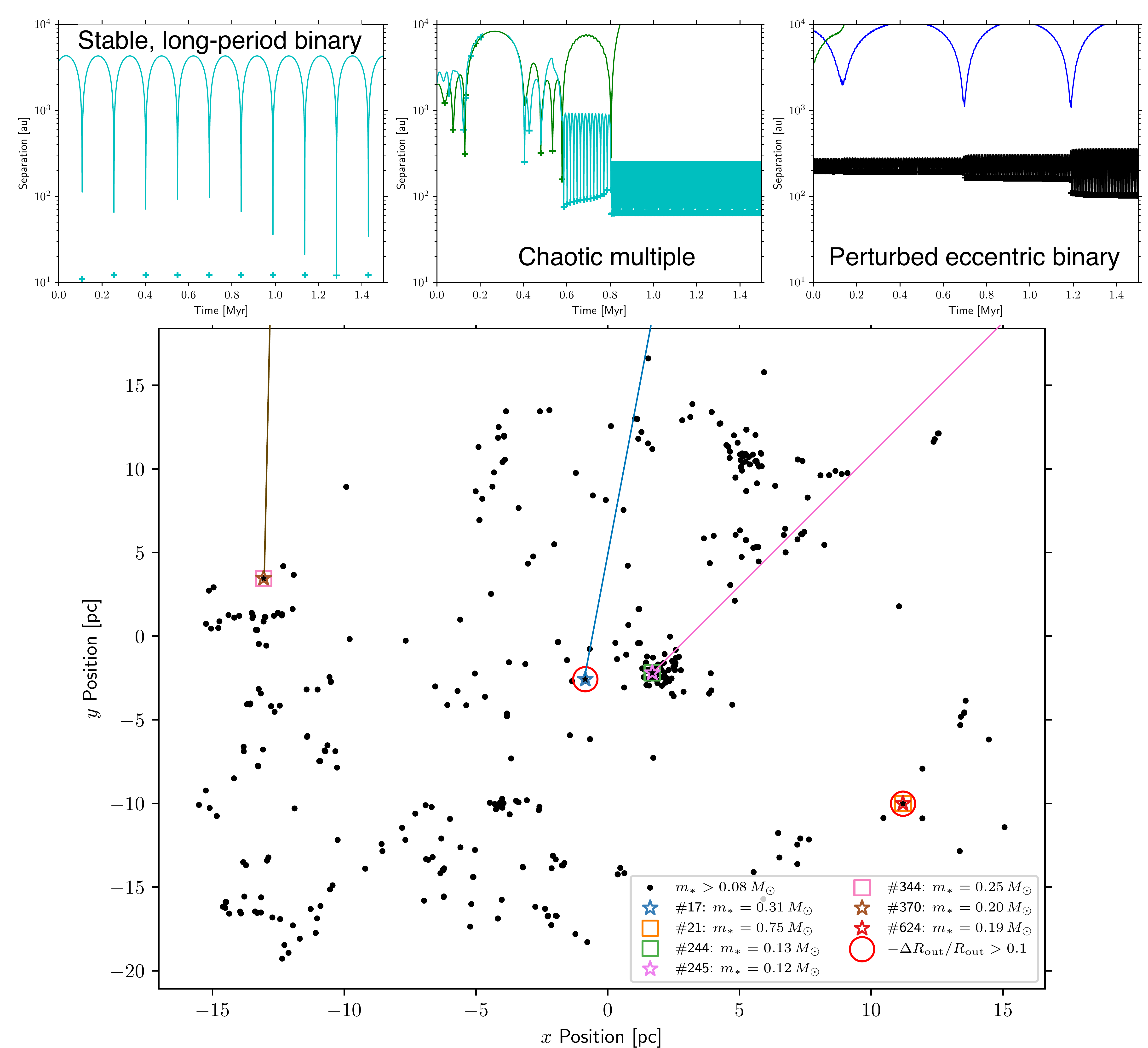}
    \caption{In the bottom panel, we show the spatial distribution of stars (mass $>0.08\, M_\odot$) at $1$~Myr and contemporary disc-truncating encounters in our simulation. We show as coloured star {or square} symbols the location of all stars that underwent encounters yielding a fractional truncation $-\Delta R_\mathrm{out}/R_\mathrm{out} > 0.01$ over the age range $1\pm 0.2$~Myr. Encounters that yielded $-\Delta R_\mathrm{out}/R_\mathrm{out} > 0.1$ are highlighted with red circles. In the panels along the top we show the separation between specific stars and their stellar neighbours over the first $1.5$~Myr of the simulation. {Each line represents the distance to a single stellar neighbour, linearly interpolated between snapshots, where neighbours that are one of the two closest stars at any given time-step are included (if they pass within $10^4$~au).} In each plot, we mark the location of a logged `closest approach' as a cross (inferred analytically from the closest time-step -- see Section~\ref{sec:encounters}). }
    \label{fig:encounter_spatial}
\end{figure*}

First, we qualitatively explore the nature of star-disc encounters in our simulation. For this we show some examples of encounters that occur at $1\pm0.2$~Myr in our simulation. The location of all these close encounters are shown in the central panel of Figure~\ref{fig:encounter_spatial}, from which we extract the following categories:

\begin{itemize}
    \item[a)] \textit{Stable, long period eccentric binary:} Possibly the most straightforward kind of encounter is a very long period binary which remains unperturbed by unbound stars, as shown in the top-left of Figure~\ref{fig:encounter_spatial}. This kind of encounter can still produce relatively large changes in outer disc radius if the orbital period is sufficiently long.
    \item[b)] \textit{Chaotic multiple:} Shown in the top-middle of Figure~\ref{fig:encounter_spatial} is an example of a bound triple system that interacts to eventually eject one of the stars at $\sim 0.8$~Myr, leaving a tighter binary that can further truncate the disc. 
    \item[c)] \textit{Perturbed eccentric binary:} If a stable binary is perturbed by an encounter with an external (unbound) star, then this may also result in a tightening of the binary. A closer periastron distance can then further truncate the disc. We show an example of this in the top-right of Figure~\ref{fig:encounter_spatial} (although in this case, the external star is in fact marginally bound).
    \item[d)] \textit{Random unbound encounters:} This is the best-studied type of encounter, the rate of which can be understood analytically at a given local stellar density and velocity dispersion. However, in our simulation, we do not find any examples of random encounters. These encounters are more common in denser regions, but in Taurus we find that the majority of star-disc interactions are mediated by wide binary companions. 
\end{itemize}

While this categorisation illuminates the value of physically motivated structure and multiplicity within models for star forming regions, they are not rigid, and types of encounter may blur into each other. Our findings emphasise that unbound encounters and binary interactions cannot be studied separately during the dynamical evolution of a young star forming region.

\subsection{Disc evolution}

\label{sec:disc_evol}
\begin{figure}
    \centering
    \includegraphics[width=\columnwidth]{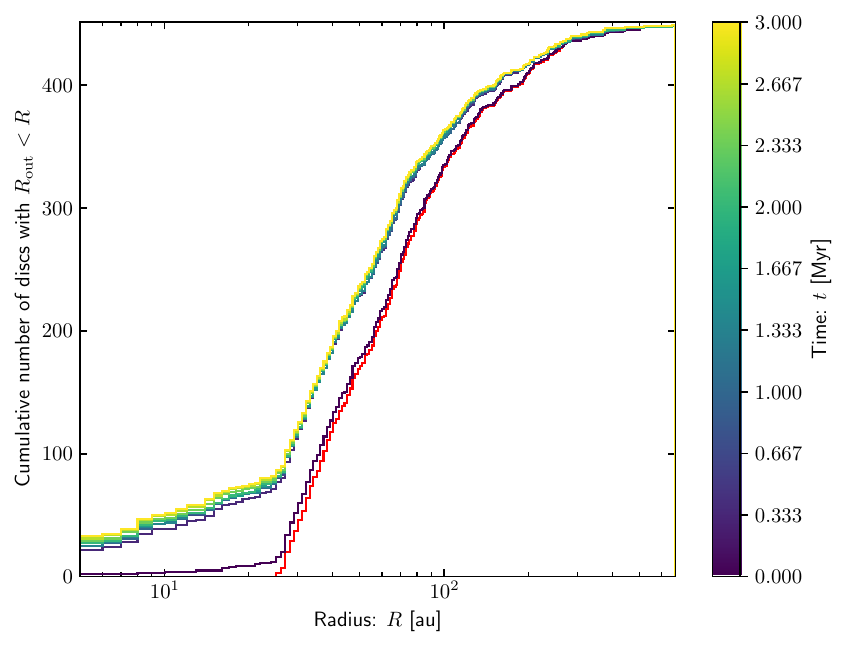}
    \caption{Disc outer radius $R_\mathrm{out}$ evolution for the population of stars with $m_*>0.08 \, M_\odot$ in our simulation. The colour bar shows the time at which the distribution is measured. The initial outer radius distribution is shown in red. }
    \label{fig:disc_r_cum}
\end{figure}

We consider how encounters shape the global disc radius distribution in Taurus. This is illustrated in Figure~\ref{fig:disc_r_cum}, where we show the cumulative distribution of disc outer radii over the course of the simulation. We find that the vast majority of discs are rapidly truncated below their initial radius by encounters, and $\sim 1/4$ are truncated below $30$~au. However, by the present time ($1$~Myr), dynamical encounters are not changing the \textit{overall distribution} of disc radii significantly. This is because, while external perturbation does sculpt the outer disc for a large fraction of the population, {disc truncation mostly occurs} in stable binaries. Observations of discs in binary systems appear broadly consistent with theoretical expectations \citep{Manara19, Rota22}.  We conclude that, for a region like Taurus, the role of encounters for the disc population as a whole is largely dominated by early interactions in binaries, rather than an ongoing process of disc truncation. 

\subsection{Truncating encounter frequency}

\label{sec:encounter_freq}
\begin{figure*}
    \centering
    \includegraphics[width=\textwidth]{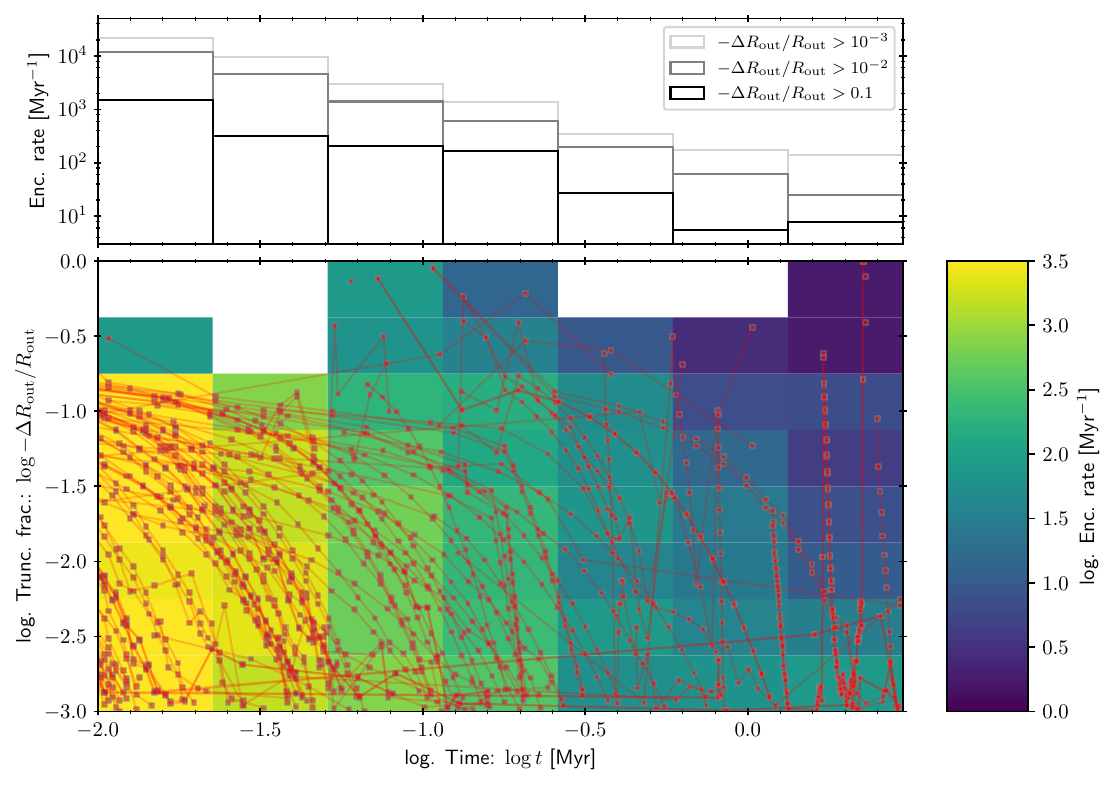}
    \caption{The global rate of disc truncating encounters for discs around stars with mass $m_*>0.08\, M_\odot$ throughout the evolution of our dynamical model. The top panel shows the rate of all truncating encounters that decrease the disc radius by at least $0.1$~percent (light gray), $1$~percent (gray) and $10$~percent (black) of the pre-encounter radius. The colour scale of the bottom panel shows the encounter rates binned into both the time of the encounter ($x$-axis) and the degree of truncation ($y$-axis). Each logged encounter is shown by a pink square, and subsequent encounters for the same disc (if any) are connected by faint pink lines. }
    \label{fig:encounter_summary}
\end{figure*}

We now turn to the primary motivation of this work, and ask the question: \textit{do we expect sufficient close encounters in Taurus to produce the examples of recent dynamical interaction?} To answer this question we require two definitions:
\begin{enumerate}
    \item Which encounters produce significant external structures (e.g. tidal tails)? 
    \item For how long does this observable external structure persist? 
\end{enumerate}

To answer these questions, we consider the rate of encounters that yield a fractional truncation $-\Delta R_\mathrm{out}/R_\mathrm{out}$ above some threshold, in the context of the observed flyby candidates in Taurus. 

\subsubsection{HV and DO Tau}

The huge extended dust bridge between HV and DO Tau \citep{Howard13} appears to be the result of a strongly truncating encounter, possibly occurring during the dynamical decay of a quadruple system \citep{Winter18c}. While the model of \citet{Winter18c} is probably not a unique scenario for producing the observed structure, we can make some quantitative arguments as to the requirements of such an encounter. Assuming some grain growth occurred within the disc, the mass of material expelled during the encounter is $M_\mathrm{ex} \sim 10^{-4} \, M_\odot$ \citep{Winter18c}, which is $\sim 10$~percent of the current mass $M_\mathrm{disc}$ of the disc around HV Tau C \citep{Stapelfeldt03}. If the surface density of the disc $\Sigma \propto R^{-1}$, for $R$ the cylindrical radius inside $R_\mathrm{out}$, this implies a fractional truncation $-\Delta R_\mathrm{out}/R_\mathrm{out}\sim M_\mathrm{ex}/M_\mathrm{disc} \sim 0.1$. To reach the present day projected separation of $\gtrsim 10^4$~au, the encounter must have occurred $\sim 0.1$~Myr ago. This would suggest a rate of $\sim 10$~encounters~per~Myr for encounters that result in a fractional truncation $-\Delta R_\mathrm{out}/R_\mathrm{out} \gtrsim 0.1$. 

\subsubsection{RW Aurigae}

For RW Aurigae, the best-fitting model explored by \citet{Dai15} had an initial outer radius of $60$~au and a final outer radius in the range $\sim 40-57$~au. This estimate is mostly based on the geometry of the spiral arm as inferred from their simplified radiative transfer. The $^{12}$CO emission detected around RW Aurigae is at least partly optically thick, so it is not possible to reliably infer a total mass that has been ejected in the encounter. The conclusions of \cite{Dai15} may change if the apparent morphology as seen in CO differs when detailed chemistry, photodissociation or radiative transfer effects are taken into account. If the physical arm is longer and wider than they assume, the authors show that a weaker encounter can be consistent with the observed structure. \citet{Dai15} also infer a time since closest approach $\sim 600$~years from their dynamical model. This conclusion may depend somewhat on the deprojected geometry of the spiral and the system orientation.

\subsubsection{UX Tau}

In the case of  UX Tau, \citet{Menard20} focused on two different flyby simulations, with outer radius $R_\mathrm{out} = 60$~au and $90$~au, both with $r_\mathrm{p} = 100$~au, with a mass-ratio of the perturber $q \sim 0.08-0.22$.  The former radius is close to the observed outer radius (post-encounter). Both simulations produce clear spiral arms and some extended structure, so a severely truncating encounter is not required. There are currently no estimates for the mass of the external structure with respect to the disc mass. From their simulations, \citet{Menard20} estimate $\sim 1000$~years since closest approach, with some margin for uncertainty in the system geometry as in the case of RW Aurigae.

\subsubsection{Overall rates}
\label{sec:overall_rates}
The considerations above lead us to conclude that a small fraction of the disc mass may be capable of producing some of the observed external structures. We therefore explore a range of fractional truncation thresholds $-\Delta R_\mathrm{out}/R_\mathrm{out}>10^{-3}$, $10^{-2}$ and $10^{-1}$. Choosing a threshold smaller than $10^{-3}$ makes little difference to the overall encounter rate. As discussed above, `weak' encounters with $-\Delta R_\mathrm{out}/R_\mathrm{out} \sim 10^{-2}$ may be sufficient to reproduce the structures observed around RW Aurigae and UX Tau, while the stronger encounter threshold would correspond to systems like HV and DO Tau. While we cannot be confident of the exact $-\Delta R_\mathrm{out}/R_\mathrm{out}$ that result in RW Aurigae and UX Tau-like systems, our results will motivate future studies quantifying the mass in the external structure surrounding systems that have undergone recent encounters, given that $M_\mathrm{ex}/M_\mathrm{disc} \sim -\Delta R_\mathrm{out}/R_\mathrm{out}$. 

The overall encounter rate is summarised in Figure~\ref{fig:encounter_summary} for these thresholds. We have binned the encounters by the time at which they occur and by their fractional disc truncation. We also show the evolution of individual discs by connected lines between individual encounters, so that the evolution of the outer disc radii in (for example) binary systems is clear. As expected from Section~\ref{sec:disc_evol}, the majority of discs are truncated rapidly in binary systems during the first $\sim 0.1$~Myr. However, there remain examples of individual encounters persisting throughout the course of the simulation. In order to explore the degree to which our results are stochastic, we run two additional versions of our experiment described in Appendix~\ref{app:stochastic}.

Quantitatively, we can see that the rate of the strongest encounters $-\Delta R_\mathrm{out}/R_\mathrm{out} > 0.1$ remains at $\sim 10$~Myr$^{-1}$ in the time range $\sim 1{-}3$~Myr (see also Appendix~\ref{app:stochastic}). Thus HV and DO Tau-like encounters, observable for $\sim 0.1$~Myr, are likely (probability $\sim 50$~percent) to be found somewhere in Taurus. We conclude that the occurrence of the HV and DO Tau encounter is expected in the context of our dynamical model. 

The more recent RW Aurigae- and UX Tau-like encounters require a considerably higher encounter rate. The rate of weaker encounters $-\Delta R_\mathrm{out}/R_\mathrm{out} > 10^{-2}$ is $\Gamma_\mathrm{enc} \sim 100$ Myr$^{-1}$, with approximately a factor order unity in stochastic variation (Appendix~\ref{app:stochastic}). This rate is a factor $\sim 2{-}3$ larger for $-\Delta R_\mathrm{out}/R_\mathrm{out} > 10^{-3}$. We can write the probability of observing at least $N$  encounters from a Poisson distribution:
\begin{equation}
\label{eq:Pobs}
    P_\mathrm{obs} (N_\mathrm{obs} \geq N) = 1- \sum_{i=0}^{N-1}\frac{ ( \Gamma_\mathrm{enc} \tau_\mathrm{obs})^{i}\exp(- \Gamma_\mathrm{enc} \tau_\mathrm{obs}) }{i !},
\end{equation}where $\tau_\mathrm{obs}$ is the period for which the encounter is observable. Given that we are equally likely to observe the disc at any stage during this period of observability, this implies the average time of observation $\langle t_\mathrm{obs} \rangle = \tau_\mathrm{obs}/2$. We therefore adopt a moderately generous $\tau_\mathrm{obs} = 2000$~years, a factor $\sim 2$ larger than the time since periastron for UX Tau and RW Aurigae. Taking a range of encounters $\Gamma_\mathrm{enc} = 100{-}200$~Myr$^{-1}$ for $N_\mathrm{obs} \geq 2$ (RW Aurigae and UX Tau) yields $P_\mathrm{obs} = 0.018 {-} 0.062$. Therefore, if encounters with $-\Delta R_\mathrm{out}/R_\mathrm{out} \gtrsim 10^{-2}$ can produce these systems, then the tension with our model are not significant (or very marginally significant at $\sim 2\, \sigma$). If a much more truncating encounter is required, then this tension may become significant. This marginal agreement underlines the importance of future studies quantifying the fraction of mass in the extended structure. 

In the absence of additional constraints, we conclude that the expected rate of encounters in Taurus is marginally sufficient to produce UX Tau and RW Aurigae without appealing to additional physics (see Section~\ref{sec:add_physics}). However, this would not be the case if the sample of known recent, truncating star-disc encounters is incomplete. Indeed, there is some reason to suspect that this may be the case, as discussed below.

\subsection{Masses of stars undergoing close encounters}
\label{sec:encrate_mstar}
\begin{figure}
    \centering
    \includegraphics[width=\columnwidth]{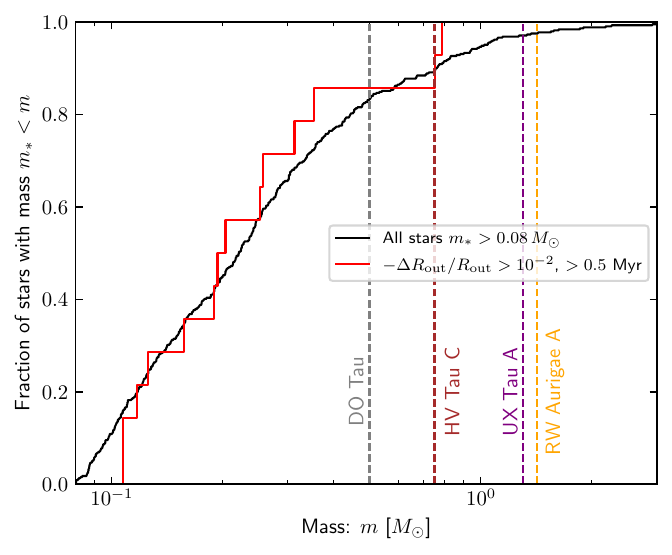}
    \caption{Cumulative distribution function for the masses $m_*$ of stars with $m_*>0.08\,M_\odot$ in our simulation (black line), compared with those that underwent a truncating encounter $-\Delta R_\mathrm{out}/R_\mathrm{out}>0.01$ (red line). The  distributions are not significantly different, with KS test probability $p_\mathrm{KS} =0.87$.  We also show mass estimates for the stars with discs that are responsible for the observed external structure. We do not show uncertainties in these estimates for clarity, but errors quoted are typically $\sim \pm 0.3 \, M_\odot$. The distribution of masses of stars that have been inferred to have experienced recent truncating encounters are significantly different from those in our simulation, with $p_\mathrm{KS} =0.023$.}
    \label{fig:mstar_enc}
\end{figure}

We can ask whether truncating encounters occur more or less frequently for high mass stars. We show the distribution of the masses of stars that undergo truncating encounters after $1$~Myr in Figure~\ref{fig:mstar_enc}. This mass distribution is indistinguishable to the overall mass function ($m_*>0.08 \, M_\odot$). By contrast, RW Aurigae A and B have masses $\sim 1.3-1.4\, M_\odot$ and  $\sim 0.7-0.9 \, M_\odot$ respectively \citep{Ghez97, Woitas01}. HV Tau C has a mass $\sim 0.5-1 \, M_\odot$ \citep{Duchene10} and DO Tau $\sim 0.3 - 0.7 \, M_\odot$ \citep{Beckwith90, Hartigan95}. UX Tau A has a mass $\sim 1.3 \, M_\odot$ and UX Tau C has mass $\sim 0.16 \, M_\odot$ \citep[][]{Kraus09, Zapata20}. The origin of the extended structures are the discs around RW Aurigae A, HV Tau C, DO Tau and UX Tau A. The masses of these stars appear to be systematically greater than what would be obtained from randomly sampling from the IMF. Assuming nominal masses of $1.4 \, M_\odot$, $0.8 \, M_\odot$, $0.5 \, M_\odot$ and  $1.3 \, M_\odot$ for RW Aurigae A, HV Tau C, DO Tau UX Tau A respectively yields KS test $p$-value $0.023$~percent when comparing to the stars that undergo encounters in our model. It is quite possible that observations could be biased to detect evidence of encounters involving more massive stars with brighter discs. However, this would imply many more undetected external structures that are evidence of encounters involving low mass stars in Taurus. Following equation~\ref{eq:Pobs}, \textit{if evidence for just one more encounter similar to RW Aurigae or UX Tau were uncovered in Taurus, this would being the $2\sigma$ encounter rate to $\Gamma_\mathrm{enc} \sim 680$~Myr$^{-1}$.} Alternatively, this would yield $P_\mathrm{obs}\approx 10^{-3}$ even for weak encounters with $\Gamma_\mathrm{enc} \sim 100$~Myr$^{-1}$. Finding further examples would introduce significant tension between the frequency of such events and the encounter rates in our model (Section~\ref{sec:overall_rates}), necessitating additional physics. This highlights the importance of future unbiased surveys to search for evidence of recent star-disc encounters in Taurus. 

We caveat our findings with the fact that we do not invoke any primordial mass segregation \citep{Zinnecker93, Moeckel10, Plunckett18}. However, mass segregation is not evident in Taurus \citep{Dib19}.

\subsection{Mechanisms to increase late encounter rates}
\label{sec:add_physics}
Given the apparently high probability that the observed sample of discs with external structure produced during a recent star-disc encounter is incomplete, we consider a number of possibilities that may enhance the frequency of encounters late-on during the dynamical evolution of Taurus. 

\subsubsection{(Viscous) re-expansion}
\label{sec:visc_exp}
For discs that experience multiple encounters that are either unbound or in long period binaries, viscous angular momentum transport \citep[e.g.][]{Lyn74} or in-fall of material from the ISM \citep[e.g.][]{Padoan05, Manara18, Kuffmeier23, Gupta23, Padoan24, Winter24} may replenish outer disc material such that subsequent encounters will have a stronger influence than under our assumption that disc radius is fixed. This replenishment should be correlated with stellar mass \citep[as suggested by observed stellar accretion rates --][]{Manara17}, so this would also preferentially enhance encounter rates for high mass stars, consistent with observations (Section~\ref{sec:encrate_mstar}). Conversely, if angular momentum is extracted by magnetohydrodynamic winds \citep[e.g.][]{Bai13} then this could exacerbate the encounter rate problem.  

\subsubsection{Self-gravity of the interstellar medium}
\label{sec:selfgrav}

\begin{figure}
    \centering
    \includegraphics[width=0.5\textwidth]{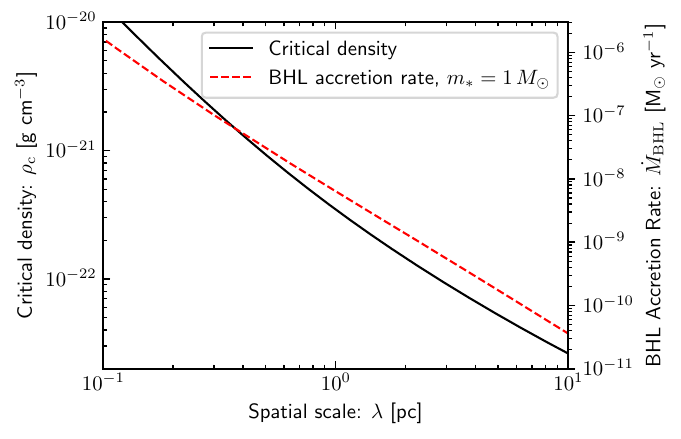}
    \caption{Critical density at which the ISM is bound against turbulent pressure (black line, left hand side $y$-axis) and the resultant Bondi-Hoyle-Lyttleton (BHL) accretion rate for a solar mass star (red dashed line, right hand side $y$-axis). Both are shown as a function of spatial scale $\lambda$.  }
    \label{fig:BHL_rate}
\end{figure}

{As discussed in Section~\ref{sec:caveats}, for practical reasons we have ignored ISM self-gravity in our calculations. If included, we would expect this self-gravity to increase the time for which groups of stars remain bound and therefore potentially enhance the frequency of truncating encounters at late times. This in-of-itself may explain any shortcoming in the frequency of encounters in our model compared to observations. However, for this to keep stars bound together, gas must be at approximately the critical density to undergo gravitational collapse. On length scales $\lambda$ relevant for the Taurus region ($\sigma_v\gg c_\mathrm{s}$ and $\lambda\ll h$, the galactic scale height), this is effectively the Jeans criterion: the gravitational potential must balance the turbulent energy. At such densities, as discussed above, a number of authors have suggested that disc replenishment by Bondi-Hoyle-Lyttleton (BHL) accretion may be substantial \citep[][]{Padoan05,Throop08, Kuffmeier23, Winter24}.}

{To test whether we expect stellar dynamics-altering ISM self-gravity to also substantially replenish protoplanetary discs, we derive the critical density $\rho_\mathrm{c}$ as a function of $\lambda$ following \citet{Winter24}, shown as the black line in Figure~\ref{fig:BHL_rate}. We then estimate the expected typical BHL accretion rate:}
\begin{equation}
    \dot{M}_\mathrm{BHL} \sim \frac{4 \pi G^2 m_*^2 \rho_\mathrm{c}}{\sigma_v^3}
\end{equation}{for a solar mass star $m_*=1\, M_\odot$, shown as the red dashed line in Figure~\ref{fig:BHL_rate}. Here $\sigma_v \propto \lambda^{0.5}$ as we assume when implementing kinematic substructure. For typical regions of size $\lambda \sim 1$~pc \citep{Schmalzl10, Joncour18}, our calculations imply a BHL accretion rate $\dot{M}_\mathrm{BHL} \sim 10^{-8} \, M_\odot$~yr$^{-1}$, comparable to observed stellar accretion rates \citep[e.g.][]{Manara23}. Therefore, if local dynamics is influenced by ISM self-gravity, we expect this also to replenish protoplanetary discs as discussed above in Section~\ref{sec:visc_exp}.} 

{We summarise that, while we cannot rule out ISM self-gravity as a possible driver of present day encounters in Taurus, this would have extremely interesting consequences for disc evolution. In particular, it would imply that replenishment from the ISM is indeed common.}

\subsubsection{Tidal torques from the disc on the perturber}

If the disc is sufficiently massive and the encounter sufficiently close, then it is possible that the tidal dissipation of the orbital energy of the perturber leads to capture and tightening of the bound orbit \citep{Clarke93, Ostriker94, Munoz15}. In this case the subsequent encounters may be stronger as the periastron separation decreases. This is a promising way to generate delayed strong encounters in binary/multiple systems where multiple encounters have occurred, particularly given the evidence for several close passages in the RW Aurigae system \citep{Rodriguez18}. Since the disc mass is a superlinear function of star mass \citep[e.g.][]{Ansdell16}, this mechanism could also be more effective for perturbations to discs around massive stars. A key question however, is whether discs retain sufficient mass to generate orbital decay to the present day; this may be the subject of future hydrodynamic simulations for RW Aurigae and UX Tau.

 \subsubsection{Embedded planets or brown dwarfs}
 
 If brown dwarfs or massive planets form and remain embedded in the protoplanetary discs, then it is possible that during the encounter they undergo large eccentricity perturbations \citep{Heg96} or chaotic evolution, perhaps to be captured by the perturbing star \citep[e.g.][]{Fregeau04}. This would effectively be a late-stage enhancement of the binary fraction, which may be expected to enhance the encounter rate. As for the previous two mechanisms, massive planets appear to be more common around high mass stars \citep{Johnson10}. While an interesting possibility in the context of spiral arms in discs, given the presence of a stellar-mass perturber in the cases of RW Aurigae and UX Tau this does not appear to be a convincing explanation. 

 \subsubsection{Spatial variation of the binary fraction}
   In our simulations, we initiated binaries independently of their location. It is plausible that binary formation is more probable in regions of enhanced local density, where enhanced tides can in principle result in fragmentation into multiple systems \citep{Horton01}. This would enhance the degree to which binaries typically interact with other stars, and may therefore lead to enhanced encounter rates. However, RW Aurigae at least does not appear to be located in a region of high stellar density \citep[e.g.][]{Pfalzner21}. 

\subsubsection{Prospects}
While these explanations discussed in this section are not yet necessary based on existing constraints, two future developments may change this conclusion. First, the quantity of mass in the extended structure surrounding RW Aurigae and UX Tau could be a substantial fraction of the disc mass (that may imply $|\Delta R_\mathrm{out}/R_\mathrm{out}|\gg 10^{-2}$). Second, any new discoveries of extended structure in Taurus originating from a star-disc encounter will considerably increase the required encounter rate to $\Gamma_\mathrm{enc} \sim 680$~Myr$^{-1}$, much greater than the rate for even weak encounters in our model $\Gamma_\mathrm{enc} \sim 200$~Myr$^{-1}$. Should an additional recent encounter be discovered, and given the arguments outlined above, possibly the most promising mechanism for enhancing the encounter rate is the (viscous) re-expansion or ISM replenishment of the disc. This hypothesis should be quantitatively investigated in the event of additional discoveries. 

\section{Conclusions}
\label{sec:conclusions}
We have presented an N-body model for the Taurus star forming region which is physically and empirically motivated.  Our initial stellar population is generated probabilistically from a turbulent gaseous medium, via a zoom-in on a Gaussian field using the cosmology simulation tool \textsc{genetIC} \citep{Stopyra20}. We applied the inferred size-velocity relation in Taurus to generate an initial velocity dispersion, combined with an empirically motivated binary population. Without any additional tuning, this yields a dynamical model with excellent agreement to the observed structure in Taurus by the metrics of the pair separation distribution and separation-velocity correlation. This has allowed us, for the first time, to accurately assess the global frequency of encounters in the Taurus star forming region.

In this way we have shown that, like the bull that is its namesake, stars in the Taurus star forming region rarely settle for one close encounter. Instead, star-disc systems act as stellar matadors, often enduring several close approaches with a neighbouring star. The closest approach distances change stochastically over time as binaries are themselves perturbed by low velocity neighbours. High order multiples can also form and break up on time-scales much shorter than the lifetime of Taurus. As a result, strong encounters at the present day can occur in one of four ways. The most common are: during the evolution of a chaotic multiple system; when an eccentric binary is perturbed; or as a result of a close approach in a very long period eccentric binary. Random encounters between single stars are rare, with close encounters in Taurus being mostly mediated through a binary companion. These categories can be blurred, and in some sense to distinguish between binary and unbound encounters is a false dichotomy. Since binaries and nearby low velocity stars influence each other, {\textit{inferring accurate encounter rates requires that the spatial and kinematic structure of a region is quantitatively taken into account. }}

Overall, $\sim 1/4$ of discs are truncated below $30$~au by dynamical encounters. However, the majority of these dynamical truncation  events happen in the first few $0.1$~Myr of the cluster evolution, over the course of a few binary periods. After this time, the role of encounters in sculpting the overall distribution of disc radii is largely finished in a low density star forming region such as Taurus.

Nonetheless, individual strong encounters still occur over the region as a whole, and we consider whether the examples of HV and DO Tau \citep{Howard13, Winter18c}, RW Aurigae \citep{Dai15, Rodriguez18} and UX Tau \citep{Zapata20, Menard20} should be observed in our model. We conclude that events resembling HV and DO Tau can occur at a rate of $\sim 10$~Myr$^{-1}$ and therefore, given that it has remained observable for $\sim 0.1$~Myr, observing one such event at any given time is expected. If weak encounters that eject only $\sim 0.1- 1$~percent of the disc mass are responsible for the external structure around RW Aurigae and UX Tau, then they are also expected given the encounter rate $\Gamma_\mathrm{enc} \sim 100{-}200$~Myr$^{-1}$ in our model. However, we highlight that the systems that have been inferred to have experienced recent truncating encounters do not appear to be consistent with random drawing from the mass function, while those in our model are. This hints that the known sample of discs that have recently experienced truncating encounters is incomplete: {\textit{just one more observational example would be in strong tension with our model, implying additional physics.}}

We discuss a number of physical mechanisms that should be explored by future work that may yield enhancements in the rate of truncating encounters, such as the tightening of binaries due to star-disc interaction \citep[e.g.][]{Munoz15} or re-expansion/replenishment in the outer regions of protoplanetary discs.{ We also show that substantial replenishment via BHL accretion must be proceeding if ISM self-gravity, which we neglect in this work, is significant for the dynamical evolution of high order multiples across the Taurus complex. }

We summarise that star-disc encounters are an important probe of disc physics. This work highlights the need for a systematic search for extended structure generated by star-disc encounters in Taurus.

\begin{acknowledgements}{We thank the anonymous referees for their useful comments that helped improve the clarity of this manuscript.} This project has received funding from the European Research Council (ERC) under the European Union’s Horizon research and innovation programme (grant agreement No. 101002188, project PROTOPLANETS, and grant agreement No. 101042275, project Stellar-MADE).
AJW has received funding from the European Union’s Horizon 2020 research and innovation programme under the Marie Skłodowska-Curie grant agreement No 101104656. LS was supported by NSFC grant Nos. 11890692, 12133008, and 12221003. LS thanks T Fang for support and acknowledges science research grants from NSFC, grant Nos. 11890692, 12133008, and 12221003. This work has made use of data from the European Space Agency (ESA) mission {\it Gaia} (\url{https://www.cosmos.esa.int/gaia}), processed by the {\it Gaia} Data Processing and Analysis Consortium (DPAC, \url{https://www.cosmos.esa.int/web/gaia/dpac/consortium}). Funding for the DPAC has been provided by national institutions, in particular the institutions participating in the {\it Gaia} Multilateral Agreement.
\end{acknowledgements}

%-------------------------------------------------------------------
\bibliographystyle{aa}
\bibliography{references}

\appendix
\section{Output frequency exploration}
\label{app:output_freq}
\begin{figure*}
    \centering
    \includegraphics[width=0.8\textwidth]{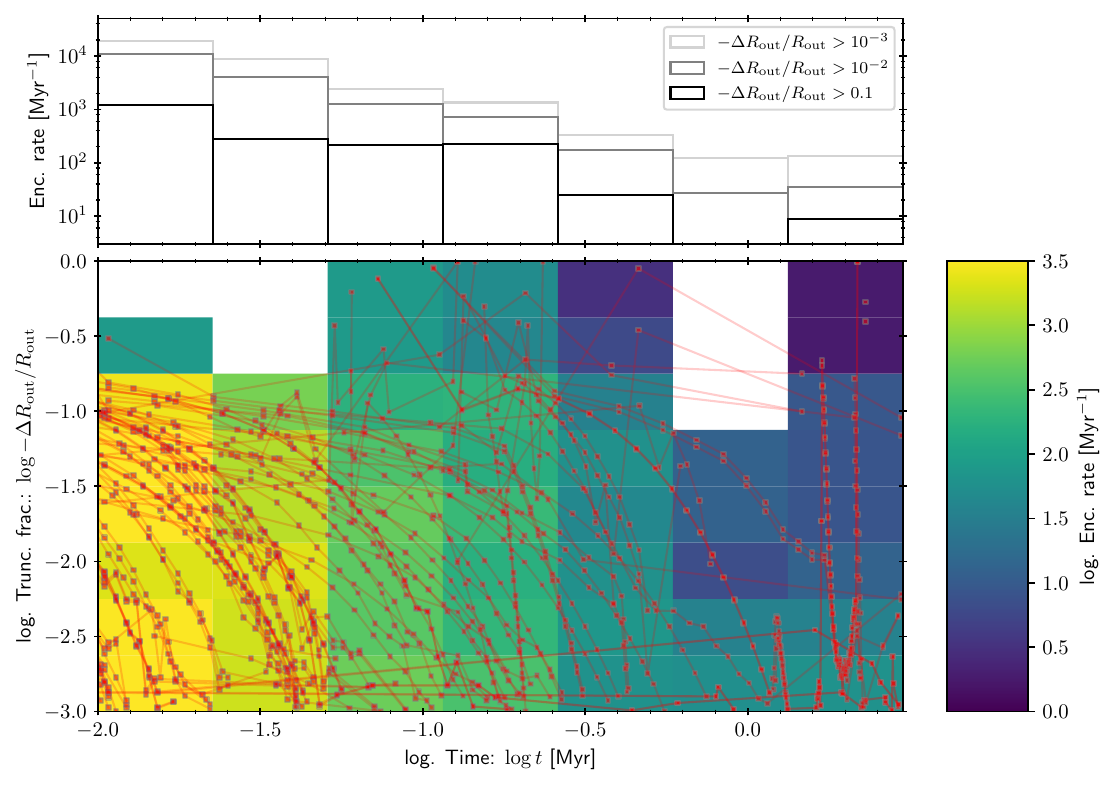}
    \caption{As in Figure~\ref{fig:encounter_summary}, but for a factor ten higher output frequency.}
    \label{fig:encsumm_hf}
\end{figure*}

\begin{figure}
    \centering
    \includegraphics[width=\linewidth]{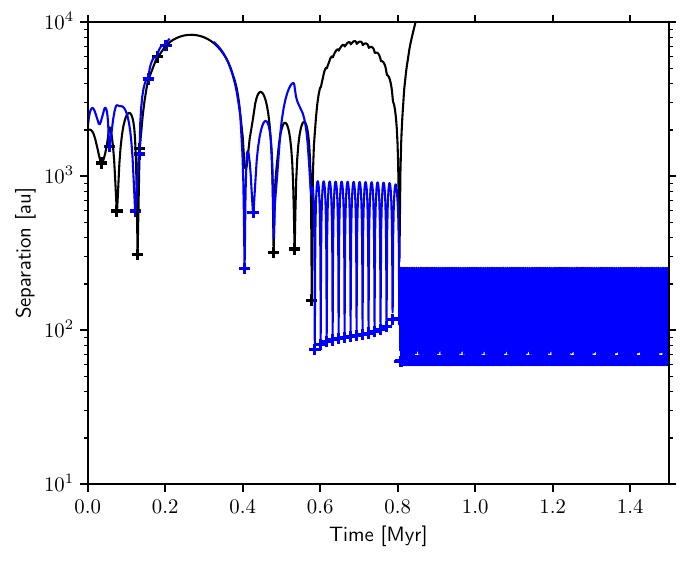}
    \caption{An example of a chaotic multiple interaction in our fiducial model. We show the separation of two neighbouring stars from a third star as a function of time. Crosses mark the location where the closest encounter is recorded by our post-processing analysis.  }
    \label{fig:chaotic_fid}
\end{figure}
\begin{figure}
    \centering
    \includegraphics[width=\linewidth]{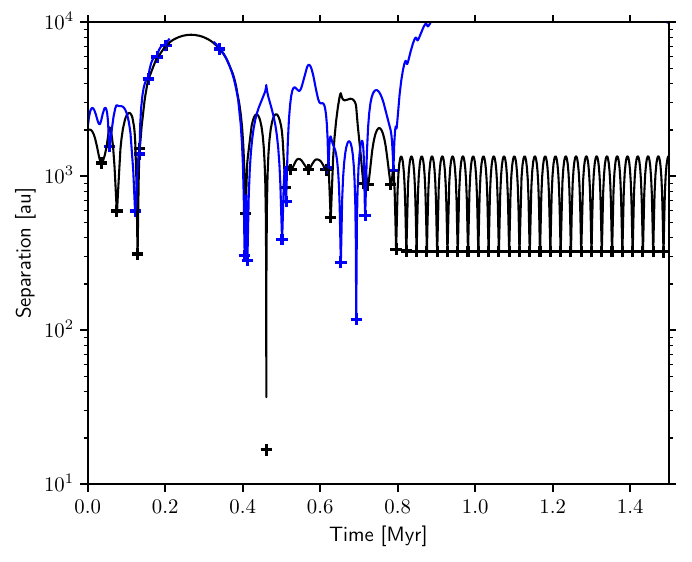}
    \caption{As in Figure~\ref{fig:chaotic_fid}, for the same stars  but with a factor ten higher output frequency.  }
    \label{fig:chaotic_hf}
\end{figure}
\begin{figure}
    \centering
    \includegraphics[width=\linewidth]{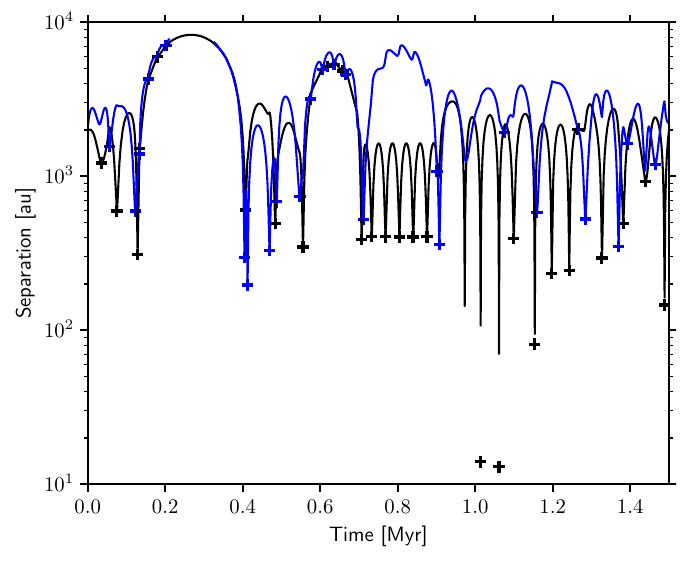}
    \caption{As in Figure~\ref{fig:chaotic_fid}, but with a factor five smaller time-step between the adjustment of simulation paramaters (\texttt{DTADJ} parameter in \textsc{Nbody6++}). }
    \label{fig:chaotic_hadj}
\end{figure}

Our approach to computing the encounter frequency has been to post-process high frequency outputs, following the method described in Section~\ref{sec:encounters}. For our fiducial model, we adopt an output frequency of one every 323 years. While this frequency is sufficient to capture even binary encounters if the semi-major axis $a \gtrsim 50$~au, it is possible that in some instances binary-single or binary-binary interactions result in miscalculation of the true encounter properties. Here we test if such events may change our results.

While it would be laborious (and challenging) to check every single star's encounter history in our sample for examples where our approach does not capture close encounters, we can more directly test our choice on the disc truncation history. We therefore decrease the output time-step by factor ten (one every $32$~years), and repeat the same analysis on the rate of disc truncation as before. The results are summarised in Figure~\ref{fig:encsumm_hf}, which can be compared directly to Figure~\ref{fig:encounter_summary}. Statistically and qualitatively, our results are very similar. We therefore conclude that our approach for extracting encounters is not substantially altering our conclusions. However, we do notice some small differences in the encounter history that correspond to a small number of individual encounter histories. These differences are not important for our conclusions, but we investigate them as follows. 

We first identify an example of a system for which we observe differences in the encounter history depending on the output frequency. We show one such encounter history in Figure~\ref{fig:chaotic_fid} for our fiducial model, which is a chaotic triple interaction. The close encounters identified by our algorithm are shown as crosses. Comparison with the high frequency output simulation (Figure~\ref{fig:chaotic_hf}) shows that at early times, both encounter histories are identical. They diverge after $\sim 0.3$~Myr, from which point the dynamics of the systems evolve chaotically to different end-states. This suggests that the differences in the encounter histories inferred from Figure~\ref{fig:encsumm_hf} compared to Figure~\ref{fig:encounter_summary} are not due to differences in our encounter-extraction algorithm, but differences in the N-body integration itself. While in principle changing the frequency of outputs should not alter the integration, \textsc{Nbody6++} performs a number of accountancy operations at the time of output. It is possible that these operations slightly alter other numbers in the code that enter into numerical calculations via, for example, the adjustment of parameters. We demonstrate that altering the parameter adjustment time-step (\texttt{DTADJ}) can have a similar influence on the dynamical evolution of chaotic multiples in Figure~\ref{fig:chaotic_hadj}, where we reduce this time-step by five compared to our fiducial model. 

We do not here investigate what changes to the output time-step lead to an altered chaotic evolution of high order multiples when using the \textsc{Nbody6++} code. By the nature of chaotic interactions, such changes may be tiny (such as machine precision) and in this case no particular solution is obviously more accurate. This is particularly irrelevant physically, since for these cases other processes may also change the dynamical outcome. However, we are satisfied that the statistical distribution of star-disc encounters our model is not dependent on our choice of output time-step, and that our algorithm for extracting encounters is adequate for our purposes.

\section{Stochastic encounter history}

\label{app:stochastic}

\begin{figure}
    \centering
    \includegraphics[width=\columnwidth]{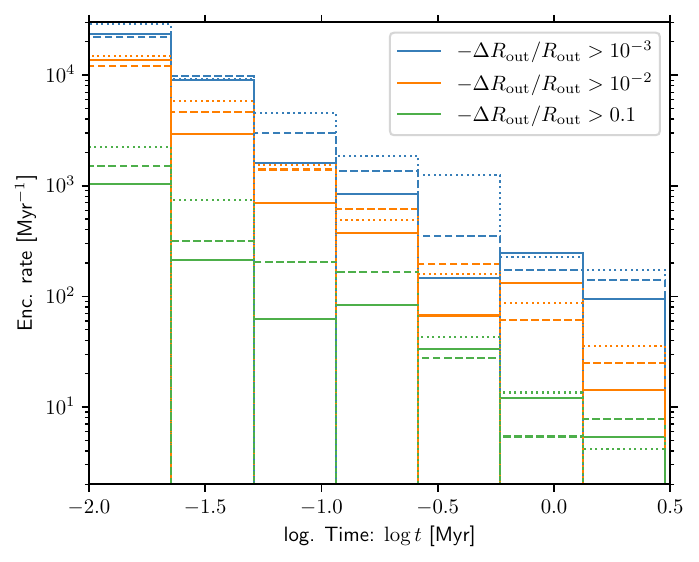}
    \caption{Stochastic variation of the encounter rate between different simulations (shown by different line styles). The colour of the lines refers to the threshold truncation used to define the encounter (as in Figure~\ref{fig:encounter_summary}).}
    \label{fig:stoch_encrate}
\end{figure}

\begin{figure*}
    \centering
    \includegraphics[width=0.8\textwidth]{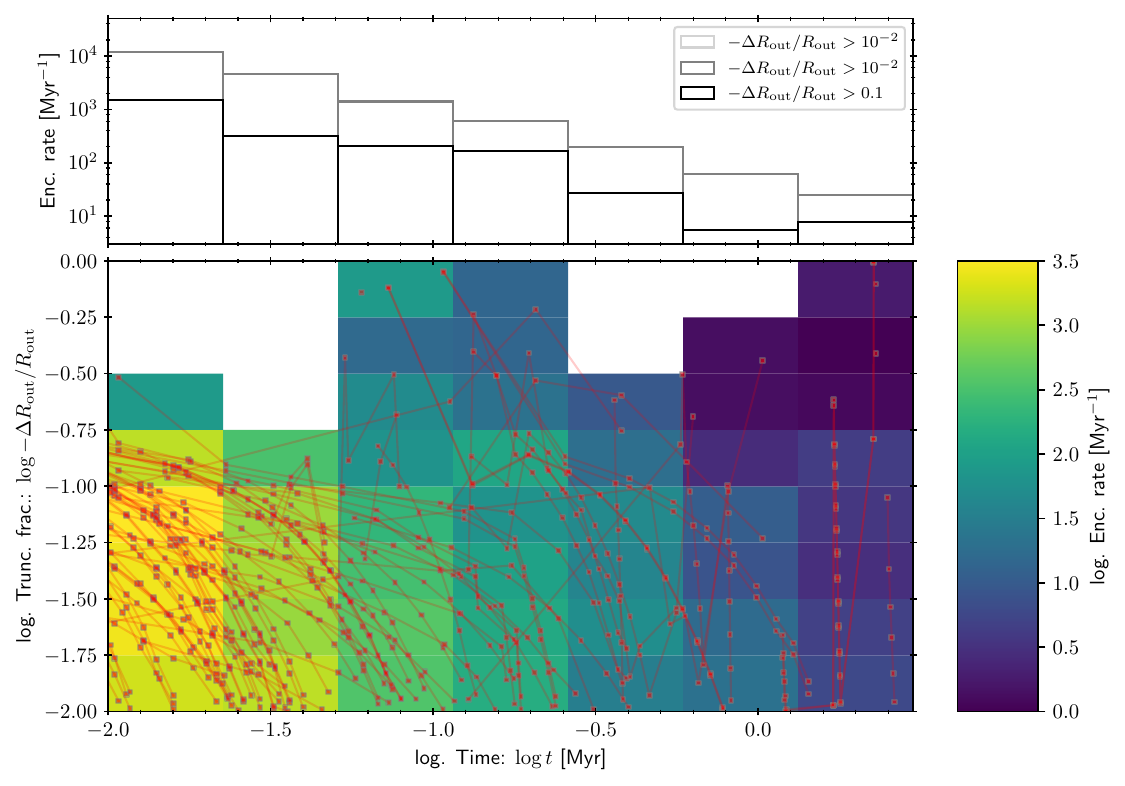}
    \caption{As in Figure~\ref{fig:encounter_summary} but for a second random drawing of the stellar and binary population. }
    \label{fig:encounter_summary_2}
\end{figure*}

\begin{figure*}
    \centering
    \includegraphics[width=0.8\textwidth]{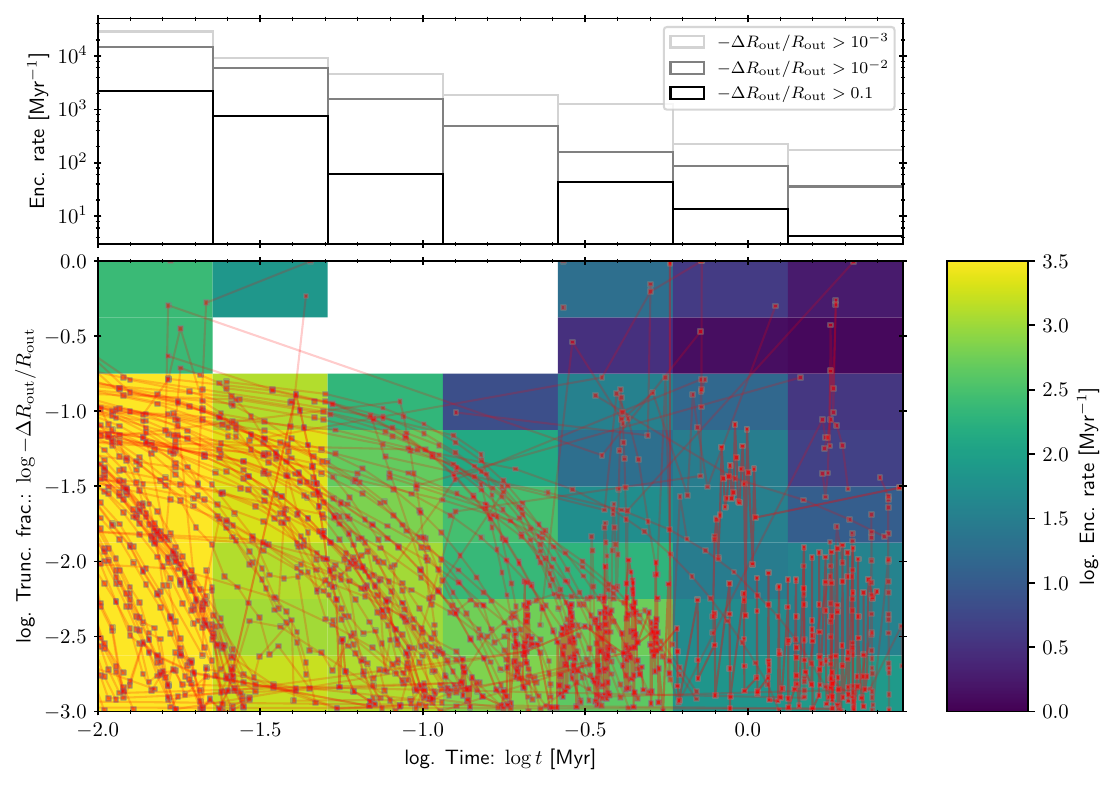}
    \caption{As in Figure~\ref{fig:encounter_summary_2} but for a third random drawing of the stellar and binary population.}
    \label{fig:encounter_summary_3}
\end{figure*}

To ensure that we are not dominated in our quantitative conclusions by stochastic variations in the encounter rates, we perform two additional numerical integrations of a Taurus-like region. We do this by performing two resampling experiments for the same density field as we adopt for our fiducial model, then adding a new binary population. The outcome comparing the encounter rates across all the simulations is shown in Figure~\ref{fig:stoch_encrate}. Figures comparable to Figure~\ref{fig:encounter_summary} are shown in Figures~\ref{fig:encounter_summary_2} and~\ref{fig:encounter_summary_3}. While we are not able to repeat the experiment enough times to gain a full distribution at each time interval, we can estimate a factor $\sim 2$ variation in the encounter rates is typical. We conclude that the rate of truncating encounters that we predict is only stochastic to within a factor of order unity. 

\end{document}